\documentclass[twocolumn]{aastex631}

\usepackage{physics}
\usepackage{graphicx}
\usepackage{amsmath,amssymb}
\usepackage{bm}
\usepackage{color}

\newcommand{\FB}{F_{\mathrm{B}}}

\newcommand{\cs}{c_{\mathrm{s}}}

\newcommand{\RBHL}{R_{\mathrm{BHL}}}

\defcitealias{1999ApJ...513..252O}{O99}

\begin{document}

\title{Gas Dynamical Friction on Accreting Objects}

\correspondingauthor{Tomoya Suzuguchi}
\email{suzuguchi@tap.scphys.kyoto-u.ac.jp}

\author[0009-0005-1459-1846]{Tomoya Suzuguchi}
\affiliation{Department of Physics, Kyoto University, Sakyo, Kyoto 606-8501, Japan}

\author[0000-0001-7842-5488]{Kazuyuki Sugimura}
\affiliation{Faculty of Science, Hokkaido University, Sapporo, Hokkaido 060-0810, Japan}


\author[0000-0003-3127-5982]{Takashi Hosokawa}
\affiliation{Department of Physics, Kyoto University, Sakyo, Kyoto 606-8501, Japan}

\author[0000-0002-8125-4509]{Tomoaki Matsumoto}
\affiliation{Faculty of Sustainability Studies, Hosei University, Fujimi, Chiyoda, Tokyo 102-0033, Japan}



\begin{abstract}
The drag force experienced by astronomical objects moving through gaseous media (gas dynamical friction) plays a crucial role in their orbital evolution. \citet{1999ApJ...513..252O} derived a formula for gas dynamical friction by linear analysis, and its validity has been confirmed through subsequent numerical simulations. However, the effect of gas accretion onto the objects on the dynamical friction is yet to be understood. In this study, we investigate the Mach number dependence of dynamical friction considering gas accretion through three-dimensional nested-grid simulations. We find that the net frictional force, determined by the sum of the gravitational force exerted by surrounding gas and momentum flux transferred by accreting gas, is independent of the resolution of simulations. Only the gas outside the Bondi-Hoyle-Lyttleton radius contributes to dynamical friction, because the gas inside this radius is eventually absorbed by the central object and returns the momentum obtained through the gravitational interaction with it. In the subsonic case, the front-back asymmetry induced by gas accretion leads to larger dynamical friction than predicted by the linear theory. Conversely, in the slightly supersonic case with the Mach number between 1 and 1.5, the nonlinear effect leads to a modification of the density distribution in a way reducing the dynamical friction compared with the linear theory.  At a higher Mach number, the modification becomes insignificant and the dynamical friction can be estimated with the linear theory. We also provide a fitting formula for dynamical friction based on our simulations, which can be used in a variety of applications.
\end{abstract}


\section{Introduction} \label{sec:intro}

Moving objects interacting with their surrounding media, such as stars and gases, are ubiquitous in the universe. Such moving objects include binary stars orbiting in common envelopes \citep{2008ApJ...672L..41R}, planets orbiting in protoplanetary disks \citep{2013MNRAS.428..658T}, and supermassive black holes (SMBHs) with $> 10^6\,M_\odot$ moving in galaxies and in some cases forming pairs with other SMBHs \citep{2000ApJ...536..663N,2005ApJ...634..921A, 2004ApJ...607..765E,2005ApJ...630..152E}.

The formation of SMBH binaries resulting from galaxy mergers has been of great interest for a long time. After two galaxies hosting their own SMBHs merge, the SMBHs are believed to reduce their distance through interactions with stars and gases until they eventually merge emitting gravitational waves (GWs) \citep{1980Natur.287..307B}. Recently, some SMBH binaries separated on the kpc scale have been observed as dual active galactic nuclei \citep[see, e.g.,][for a review]{2019NewAR..8601525D}. Moreover, NANOGrav has presented compelling evidence of a stochastic GW background consistent with SMBH binary mergers \citep{2023ApJ...951L...8A}.

Recent discoveries of SMBHs at redshifts greater than six have prompted investigations into their origins in the early universe \citep[see, e.g.,][for a review]{2020ARA&A..58...27I}. While various SMBH formation scenarios have been proposed, most of them presume that SMBHs have grown from smaller BHs, predicting a large number of intermediate-mass black holes (IMBHs) with $\lesssim 10^6\,M_\odot$ harbored in primeval galaxies. Consequently, mergers of such galaxies lead to formation of IMBH binaries, whose GWs emitted at their mergers are a major target for space interferometers such as Laser Interferometer Space Antenna (LISA, \citealt{2017arXiv170200786A,2023LRR....26....2A}), TianQin \citep{2016CQGra..33c5010L}, and Taiji \citep{2020IJMPA..3550075R}.

To estimate the event rate of the GWs from IMBH binaries, understanding their cosmological evolution is crucial. However, it is often impractical to follow the evolution of IMBH binaries in cosmological simulations resolving pc or sub-pc scale dynamics, where they interact with surrounding stars and gases. Therefore, it is necessary to introduce an appropriate sub-grid model that can describe the effect
of such interactions on the motion of IMBH binaries in cosmological simulations \cite[e.g.][]{2019MNRAS.483.3488B}. The interactions of a moving object with its surrounding medium have been studied for a long time. \citet{1943ApJ....97..255C} explored the concept of dynamical friction, which is a resistive force that acts on an object moving in a stellar distribution. He developed an analytical formula to describe this force by considering the interactions between the object and the stars as a two-body gravitational problem. Later, \citet{1999ApJ...513..252O} (hereafter \citetalias{1999ApJ...513..252O}) studied the gas dynamical friction, which is the resistive force resulting from the movement of an object through a surrounding gaseous medium \citep[see also][]{1990A&A...232..447J}. She examined the linear density perturbations caused by the object and derived a formula for the frictional force by integrating the gravitational force from the surrounding gas. 

The analytical formula proposed by \citetalias{1999ApJ...513..252O} has been extensively tested through numerical simulations, and it has been confirmed that the formula accurately predicts the resistive force in both linear and circular motion cases, as long as the moving object only causes small density fluctuations, for which linear analysis is applicable \citep[e.g.][]{1999ApJ...522L..35S, 2001MNRAS.322...67S,2007ApJ...665..432K, 2004ApJ...607..765E,2008ApJ...679L..33K}. However, when a density perturbation of order unity appears in the vicinity of the object, the formula obtained by the linear analysis deviates from the actual friction force in the nonlinear regime \citep[e.g.][]{2009ApJ...703.1278K,2016A&A...589A..10T}. It is noteworthy that these previous simulations did not take into account the accretion of gas. Further numerical research is needed to gain a better understanding of the effects of gas accretion on the gas dynamical friction. 

Gas accretion onto moving objects has been studied through analytic consideration \citep[e.g.][]{1939PCPS...35..405H,1944MNRAS.104..273B} and hydrodynamic simulations (e.g., \citealt{1971MNRAS.154..141H,1979MNRAS.188...83H,1985MNRAS.217..367S,1994A&AS..106..505R,1994ApJ...427..342R,1994ApJ...427..351R,1995A&AS..113..133R,1996A&A...311..817R,2004NewAR..48..843E} for a review) for the case of uniform linear motion in a homogeneous medium. The case in which the central object is a BH and exerts radiative feedback on the surrounding gas has also been studied through radiative hydrodynamic simulations \citep[e.g.][]{2013ApJ...767..163P,2017ApJ...838..103P,2020MNRAS.496.1909T,2020MNRAS.495.2966S, 2021PASJ...73..929O,2024MNRAS.528.2588O}. However, the focus of the above studies is mainly on mass accretion growth rather than the gas dynamical friction\footnote{Exceptionally, \cite{2017ApJ...838..103P} and \cite{2020MNRAS.496.1909T} studied the effect of the gravitational pull by the shell formed on the upstream side of the object due to radiative feedback.}. Understanding of the latter is still limited even in the simplest setup of linear motion in homogeneous medium without any feedback.  While \cite{1985MNRAS.217..367S} and \cite{1995A&AS..113..133R, 1996A&A...311..817R} have obtained the resistive force acting on the objects, taking into account the effects of gas accretion, the force apparently depends on the resolution, and the numerical convergence of dynamical friction has not been addressed. Furthermore, their simulations were limited to a few specific Mach numbers, and the Mach number dependence of dynamical friction has not been fully determined. Therefore, a comprehensive comparison with Ostriker's formula is yet to be done in cases with gas accretion. 

In this paper, we perform three-dimensional numerical simulations to investigate the dynamical friction exerted on an object moving at a constant velocity in a homogeneous gaseous medium while accreting the surrounding gas. We calculate the dynamical friction as a net force due to the gravitational pull of surrounding gas and momentum flux of accreting gas. We find that the net force converges with increasing spatial resolution, as the resolution dependences of the gravitational pull and momentum flux cancel out with each other. Furthermore, through simulations with varying Mach numbers, we obtain the dynamical friction as a function of the Mach number, including the full nonlinear effects with accretion, and give a physical interpretation by comparing our results with Ostriker's linear theory. We provide a dynamical friction formula based on our results, which is an updated version of Ostriker's formula and can serve as a more realistic subgrid model in future cosmological simulations in which the gas dynamical friction cannot be directly calculated. 

The rest of the paper is organized as follows. Section \ref{sec:theory} provides a brief overview of the theoretical basis, including the linear perturbation theory of the gas dynamical friction given by \citetalias{1999ApJ...513..252O}. Section \ref{sec:method} outlines our numerical simulations and the cases examined. Section \ref{sec:result} presents the results of the numerical simulations and offers a physical interpretation of the results. Section \ref{sec:discussion} describes the application of our friction formula and compares our results with those of previous studies. Finally, Section \ref{sec:conclusion} summarizes the paper.

\section{Theoretical background} \label{sec:theory}

Here, we provide a brief overview of the theoretical background that is useful in interpreting our simulation results. 

\subsection{Characteristic length scales} \label{subsec:accretion}

Consider a situation in which a static object accretes the surrounding gas in a homogeneous medium. We assume the polytropic equation of state with an index $\gamma$ for the gas.  In this situation, we can obtain a solution for transonic accretion flow with spherical symmetry 
\citep{1952MNRAS.112..195B}.
The characteristic length scale of the flow, known as the Bondi radius, is given by 
 \begin{align}
    R_{\mathrm{B}} = \frac{GM}{\cs^2},
    \label{eq:Bondi_radius}
\end{align}
where $M$ and $c_{\mathrm{s}}$ is mass of the object and sound speed, respectively.  At a radius sufficiently smaller than $R_{\mathrm{B}}$, the accretion flow can be approximated with free fall; the gas density and inward velocity follow $\rho\propto r^{-3/2}$ and $u\propto r^{-2}$, respectively. The accretion rate is given by 
\begin{align}
    \label{eq:Bondi_rate}
    \dot{M}_{\mathrm{B}} = \frac{4\pi\lambda(\gamma)(GM)^2\rho_{\infty}}{\cs^3},
\end{align}
where $\rho_{\infty}$ is the gas density at infinity, and $\lambda(\gamma)$ is a factor of order unity that depends on $\gamma$ \citep[e.g.][]{1983bhwd.book.....S}. For example, in the isothermal case ($\gamma=1$), $\lambda=e^{3/2}/4$. 

Next, consider a case in which the object has a relative velocity with respect to the surrounding homogeneous medium, and the accretion flow becomes axisymmetric. If the relative velocity is highly supersonic ($v\gg\cs$), the gas with the impact parameter smaller than the Hoyle-Lyttleton radius \citep{1939PCPS...35..405H,1944MNRAS.104..273B}, 
\begin{align} 
\label{eq:HL}
R_{\mathrm{HL}} = \frac{2GM}{v^2}, 
\end{align} 
is accreted onto the object. When the relative velocity is comparable to the sound speed ($v \sim \cs$), the typical length scale is often expressed with the Bondi-Hoyle-Lyttleton (BHL) radius,
\begin{align} 
R_{\mathrm{BHL}} = \frac{GM}{\cs^2+v^2}, 
\end{align} 
which is an interpolation of the Bondi and Hoyle-Lyttleton radii.

\subsection{Gas dynamical friction in the linear perturbation theory} \label{sec:Ostriker}

In this subsection, we briefly review the gas dynamical friction formula derived by \citetalias{1999ApJ...513..252O}. The formula is based on linear analysis and assumes a gravitational source with mass $M$ moving at a constant velocity $v$ that appears at time $t=0$ in a homogeneous medium with density $\rho_{\infty}$. The gas is assumed to follow the isothermal equation of state with $\cs$. Under these assumptions, the time evolution of linear density perturbations excited by the moving object is analytically solved. The integration of the gravitational force due to the density perturbation yields the dynamical friction, $F_{\mathrm{DF}}$, as a function of the Mach number $\mathcal{M}\equiv v/\cs$.

The resulting formula is written as
\begin{align}
    F_{\mathrm{DF}} = -F_{\mathrm{B}}\mathcal{I}_{\mathrm{lin}}(\mathcal{M}), 
    \label{eq:DF_lin}
\end{align}
with a $\mathcal{M}$-dependent factor 
\begin{equation}
\label{eq:Ostriker}
    \begin{split}
        \mathcal{I}_{\mathrm{lin}}(\mathcal{M}) = 
        \begin{cases}
            \displaystyle
            \frac{1}{\mathcal{M}^2} \Biggl[ 
            \frac{1}{2}\ln{\qty\bigg(\frac{1+\mathcal{M}}{1-\mathcal{M}})} - \mathcal{M}
            \Biggr],
            & \mathcal{M}<1, \\
            \\
            \displaystyle
            \frac{1}{\mathcal{M}^2} \Biggl[
            \frac{1}{2}\ln{\bigg( 1-\frac{1}{\mathcal{M}^2} \bigg)}
            + \ln{\Lambda}
            \Biggr], 
            & \mathcal{M}>1,
        \end{cases}
    \end{split}
\end{equation}
the Coulomb logarithm 
\begin{align}
    \ln\Lambda = \ln\left(\frac{\mathcal{M}c_{\mathrm{s}}t}{r_{\mathrm{min}}}\right), 
\end{align}
and a normalization factor with the size of radial momentum flux in the spherical (Bondi) accretion 
\begin{align}
    \label{eq:Bondi_flux}
    F_{\mathrm{B}} = \frac{4\pi(GM)^2\rho_{\infty}}{\cs^2}. 
\end{align}
The radius $r_{\mathrm{min}}$ denotes the minimum radius of the gas that contributes to the dynamical friction. In the supersonic case ($\mathcal{M}>1$), $r_{\mathrm{min}}$ was introduced as an artificial lower limit for the radial integration of the gravitational force to avoid the divergence in the integration \citepalias{1999ApJ...513..252O} and cannot be determined within the framework of linear theory. Later in this paper, we determine $r_{\mathrm{min}}$ based on our simulation results, such that it reproduces our numerical results according to Equation~\eqref{eq:Ostriker}.

Note that the above formula depends on $t$ when $\mathcal{M} > 1$, but not when $\mathcal{M} < 1$. In the linear theory, the frictional force is roughly proportional to $\ln{(r_{\mathrm{max}}/r_{\mathrm{min}})}$, where $r_{\mathrm{max}}$ is the maximum radius of gas contributing to the dynamical friction.  In the subsonic case, the radial extent to which the density perturbation propagates gives $r_{\mathrm{max}}\simeq(c_{\mathrm{s}}+v)t$, while the radius within which the density distribution maintains a front-back symmetry gives $r_{\mathrm{min}}\simeq(c_{\mathrm{s}}-v)t$. Therefore, the frictional force becomes time independent with the time dependence in $\ln{(r_{\mathrm{max}}/r_{\mathrm{min}})}$ canceling out. However, in the supersonic case, $r_{\mathrm{max}}$ is again $\simeq(c_{\mathrm{s}}+v)t$ but $r_{\mathrm{min}}$ does not have a time dependence. Therefore, in contrast to the subsonic case, the time dependence in $\ln{(r_{\mathrm{max}}/r_{\mathrm{min}})}$ does not cancel out, and the dynamical friction becomes time dependent.

Finally, we make three remarks regarding the linear formula given by Equation~\eqref{eq:DF_lin}. Firstly, the formula is not valid near $\mathcal{M} = 1$, as Equation~\eqref{eq:DF_lin} diverges logarithmically at $\mathcal{M} = 1$. Furthermore, in the supersonic case, the formula is derived under the assumption that $\mathcal{M}>1+r_{\mathrm{min}}/(c_{\mathrm{s}}t)$  \citepalias{1999ApJ...513..252O}. Secondly, when gas accretion is present, the linear approximation is accurate only in regions far away from the object ($r \gg R_{\mathrm{BHL}}$) where the density perturbation is small. Lastly, the formula does not take into account the effect of accretion.

\section{Methods} \label{sec:method}

\begin{table*}[tb]
  \caption{Cases considered}
  \label{table:parameter}
  \begin{center}
  \scalebox{1.0}{
  \begin{tabular}{cccccc}
    \hline
    $\mathcal{M}$ & $\RBHL/R_{\mathrm{B}}$ & $R_{\mathrm{box}}/R_{\mathrm{B}}$ & $R_{\mathrm{sink}}/R_{\mathrm{B}}$ & $L_{\mathrm{max}}$ & $t_{\mathrm{final}}\,[\mathrm{yr}]$  \\
    \hline
    $0.19$ & $0.97$ & $7.3\times10^1$ & $0.14$ & $5$ & $5\times10^4$  \\
    $0.37$ & $0.88$ & $7.3\times10^1$ & $0.14$ & $5$ & $6\times10^4$  \\
    $0.62$ & $0.72$ & $7.3\times10^1$ & $0.14$ & $5$ & $1\times10^5$  \\
    $0.87$ & $0.57$ & $2.9\times10^2$ & $0.14$ & $7$ & $3\times10^5$  \\
    $0.99$ & $0.50$ & $2.3\times10^3$ & $0.14$ & $10$ & $8\times10^5$  \\
    $1.05$ & $0.47$ & $2.3\times10^3$ & $0.14$ & $10$ & $3\times10^6$  \\
    $1.11$ & $0.45$ & $5.9\times10^2$ & $0.14$ & $8$ & $3\times10^6$  \\
    $1.29$ & $0.39$ & $5.9\times10^2$ & $0.14$ & $8$ & $4\times10^5$  \\
    $1.49$ & $0.31$ & $5.9\times10^2$ & $0.14$ & $8$ & $1\times10^5$  \\
    $1.73$ & $0.25$ & $5.9\times10^2$ & $0.14$ & $8$ & $5\times10^4$  \\
    $2.23$ & $0.17$ & $5.9\times10^2$ & $0.072$ & $9$ & $\geq1\times10^4$  \\
    $2.73$ & $0.12$ & $5.9\times10^2$ & $0.036$ & $10$ & $\geq1\times10^4$  \\
    \hline
  \end{tabular}
  }
\end{center}
\tablecomments{
Col.1: Mach number of the relative velocity, Col.2: ratio of the Bondi-Holye-Littleton radius to the Bondi radius, Col.3: size of the computational domain, Col.4: radius of the sink particle, Col.5: highest AMR refinement level, Col.6: time by which the accretion rate converges to a nearly constant value. In all cases, we set the mass of the object to $M=10^3\,M_{\odot}$, the initial gas density to $n_0=10^5\,\mathrm{cm}^{-3}$, and the gas temperature to $T_{0}=10^4\,\mathrm{K}$.}
\end{table*}

We perform three-dimensional hydrodynamic simulations using SFUMATO-RT \citep{2020ApJ...892L..14S,2023ApJ...959...17S,2023ApJ...950..184K,2021MNRAS.505.4197S,2023MNRAS.519.3076S}, an extension of the Adaptive Mesh Refinement (AMR) self-gravitational magneto-hydrodynamic (MHD) code, SFUMATO \citep{2007PASJ...59..905M,2015ApJ...801...77M}. The latest version of SFUMATO-RT includes adaptive ray tracing coupled with primordial chemistry \citep{2020ApJ...892L..14S,2023ApJ...959...17S}, non-ideal MHD effects \citep{2021MNRAS.505.4197S, 2023MNRAS.519.3076S}, and moment-based radiation transfer capable of handling extremely optically thick regions \citep{2023ApJ...950..184K}. While SFUMATO and SFUMATO-RT possess diverse functionalities, we only employ those related to hydrodynamics and AMR for this study.

\subsection{Simulation setup} \label{subsec:setting}

We consider a body moving linearly in a uniform isothermal gaseous medium. We use the Cartesian coordinates $(x, y, z)$, with $-x$ being the direction of motion, and the velocity of the body can be written as $\bm{v} = -v\bm{e}_{x}$, where $\bm{e}_{x}$ denotes the unit vector in the $x$ direction. In our simulations, we adopt the frame of reference of the body, i.e., we set a uniform initial velocity $\bm{u}(\bm{x},t=0) = -\bm{v}$ to the gas with the position of the body fixed, where $\bm{u}(\bm{x},t)$ is the gas velocity at the position $\bm{x}$ and the time $t$. Once a simulation starts, the gravitational force of the body begins to pull surrounding gas towards the body. The body is treated as a sink particle that accretes gas within the radius $R_{\mathrm{sink}}$ when the density exceeds the critical value $n_{\mathrm{cr}}$. The gravitational softening is effective only within the sink radius \citep{2015ApJ...801...77M}.

Our fiducial parameter set corresponds to the case of an IMBH moving through a dense medium, a situation possibly realized in the early universe (e.g., \citealt{2009ApJ...696L.146M,2009ApJ...698..766M,2012ApJ...747....9P,2013ApJ...767..163P,2016MNRAS.459.3738I,2017MNRAS.469...62S, 2022ApJ...927..237I, 2018MNRAS.478.3961S,2018MNRAS.476..673T,2019MNRAS.483.2031T,2020MNRAS.496.1909T,2021ApJ...907...74T}; see also \citealt{2020ARA&A..58...27I} for a review). However, our results are applicable to a wide variety of cases as long as the radius of objects is sufficiently smaller than Bondi radius (see Equation~\ref{eq:Bondi_radius}) because the hydrodynamic equations we solve are essentially scale-free. We set the initial gas density to $n_0 = n_{\infty} = 10^5\,\mathrm{cm}^{-3}$ and the critical density for the sink accretion to $n_{\mathrm{cr}}=(1/50)n_0$. As long as the simulations resolve the supersonic free-falling part of the accretion flows, i.e., $R_{\mathrm{sink}} \ll R_{\mathrm{BHL}}$, the value of $n_{\mathrm{cr}}$ does not affect the flow structures outside $R_{\mathrm{sink}}$.\footnote{In fact, we performed additional simulations with $n_{\mathrm{cr}}=5~n_{0}$ and $10~n_{0}$, confirming that the flow structure, accretion rate, and the dynamical friction do not depend on these different choices of $n_{\mathrm{cr}}$.} Considering the warm neutral medium, we set the gas temperature to $T_0 = 10^4\,\mathrm{K}$ and the mean molecular weight to $\mu\simeq1.26$. These values give the gas sound speed
\begin{align}
    \cs = 8.07\,\mathrm{km/s}\, \bigg(\frac{T_0}{10^4\,\mathrm{K}}\bigg)^{1/2} \bigg(\frac{\mu}{1.26}\bigg)^{-1/2}.
\end{align}
The mass of the object is set to $M = 10^3M_{\odot}$, and the corresponding Bondi radius is
\begin{align}
    R_{\mathrm{B}} = 1.36\times10^4\,\mathrm{au}\, \bigg(\frac{M}{10^3M_{\odot}}\bigg) \bigg(\frac{T_0}{10^4\,\mathrm{K}}\bigg)^{-1} \bigg(\frac{\mu}{1.26}\bigg).
\end{align}
Note that the increase in the body's mass due to accretion during a typical simulation period of $t=10^6\,\mathrm{yr}$ (see Section~\ref{subsec:case}) is approximately $\Delta M \sim \dot{M}_{\mathrm{B}}t\sim2\times10^3\,M_{\odot}$ and comparable to the body's mass. However, we neglect the mass increment of the body in this work, in order to develop an understanding of gas dynamical friction in a steady state.

\subsection{Cases considered} \label{subsec:case}

We explore a variety of cases by changing the Mach number $\mathcal{M}$ of the moving IMBH, including both subsonic and supersonic cases. The details of these cases are summarized in Table \ref{table:parameter}. We continue a simulation run until the accretion rate becomes nearly constant. The final time $t_{\mathrm{final}}$ is listed in Table \ref{table:parameter}. 

With the AMR module in the code, we set nested grids in the computational box, placing higher resolution grids near the central object. The number of the AMR base grids is set to $(N_x,N_y,N_z)=(128,128,128)$ for all the cases. We take the larger size of the computational domain $R_{\mathrm{box}}$ (defined as half the side length of computational box) with the larger Mach number for the subsonic cases. This is because, when the Mach number is higher, it takes more time for the accretion rate to reach a steady state, which is determined by $R_{\mathrm{B}}/|v-\cs|$ for $\mathcal{M} < 1$. The sink radius is chosen to sufficiently resolve the BHL radius in each calculation. Since the BHL radius becomes smaller as the Mach number increases, we set $R_{\mathrm{sink}}\simeq1.95\times10^3\,\mathrm{au}$ for $\mathcal{M} < 2.0$, and take values that are half and a quarter of this value for $\mathcal{M} = 2.23$ and $2.73$, respectively. The minimum grid size in each calculation, $R_{\mathrm{grid}}$, is set to be one-fourth of the sink radius. The highest level of AMR, $L_{\mathrm{max}}$, is determined by $R_{\mathrm{box}}=2^{L_{\mathrm{max}}-1}N_{i}R_{\mathrm{grid}}\,(i=x,y,z)$.

\subsection{Evaluation of dynamical friction}

We evaluate the dynamical friction exerted on the moving body in the following manner.  From a simulation snapshot, we calculate the gravitational pull of surrounding gas in the direction of the body's motion as
\begin{align}
    F_{\mathrm{grav}} = 
    \Biggl[ \int_V \dd^3\bm{r}\frac{\rho(\bm{r})GM}{r^2}\frac{\bm{r}}{r} \Biggr] \cdot\frac{\bm{v}}{v},
\label{eq:DF}
\end{align}
where $V$ is the entire computational domain excluding the interior of the sink particle.
Next, we compute the momentum flux transferred by accreting gas in the direction of the object's motions
\begin{align}
    F_{\mathrm{acc}} = 
    \Biggl[ \int_{S_{\mathrm{sink}}} \dd\bm{S}\cdot\rho(\bm{r})\bm{u}\bm{u} \Biggr] \cdot\frac{\bm{v}}{v},
\label{eq:momentum}
\end{align}
where $S_{\mathrm{sink}}$ is the surface of the sink particle.
We then obtain the net frictional force, which we simply call dynamical friction throughout this paper, as the sum of these quantities, 
\begin{align}
F_{\mathrm{DF}} := F_{\mathrm{grav}} + F_{\mathrm{acc}}.
\label{eq:net}
\end{align}
According to the above definition, the body is decelerated when $F_{\mathrm{DF}} < 0$ and accelerated when $F_{\mathrm{DF}} > 0$.

\section{Results} \label{sec:result}

\subsection{Representative cases with $\mathcal{M}\simeq 0.6$ and $1.5$}

Here we focus specifically on the cases with $\mathcal{M}\simeq0.6$ and $1.5$ as representative subsonic and supersonic cases, respectively. We analyze the time evolution of the accretion rate and friction, paying attention to numerical convergence with improving spatial resolution.

\subsubsection{Snapshots of density perturbations}

\begin{figure*}[htb]
  \centering
  \includegraphics[keepaspectratio, scale=0.55]{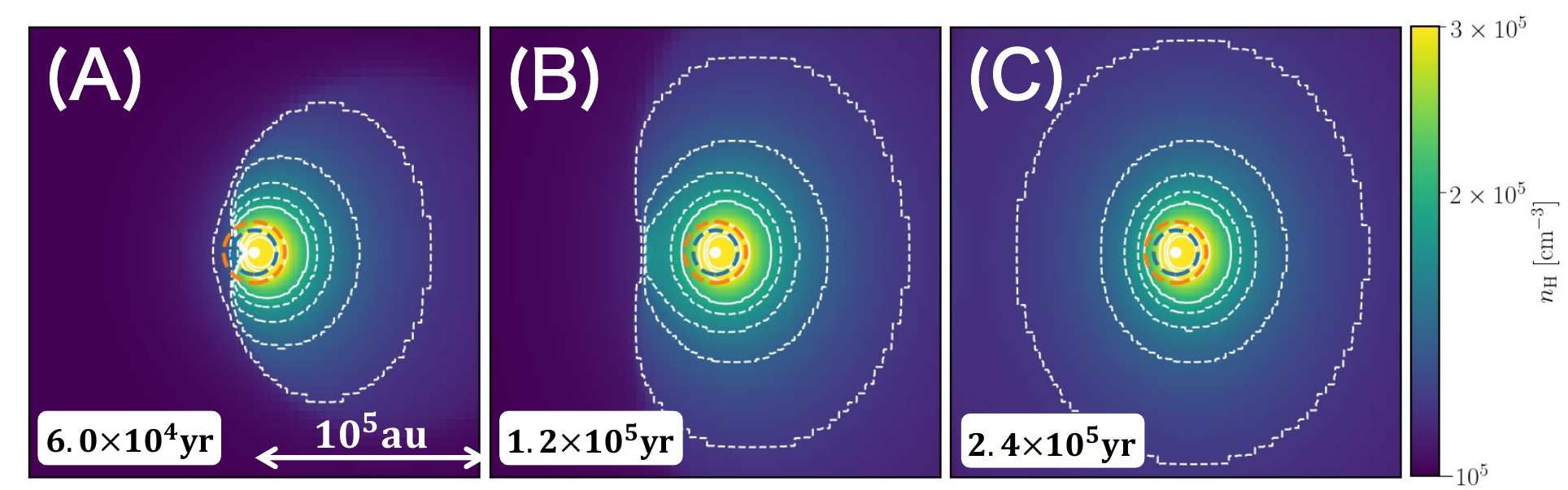}
  \caption{
  The evolution of density perturbations caused by a body moving at a subsonic speed of $\mathcal{M}\simeq0.62$. Panels (A), (B), and (C) represent the time sequence, corresponding to epochs at (A) $6\times10^4\,\mathrm{yr}$, (B) $1.2\times10^5\,\mathrm{yr}$ and (C) $2.4\times10^5\,\mathrm{yr}$, respectively. The body is assumed to move horizontally to the left.  The white filled circle at the center illustrates the sink particle. The solid lines represent isodensity contours, with the closest to the center denoting $5 \times10^5\,\mathrm{cm}^{-3}$ and the others spaced at intervals of $10^5\,\mathrm{cm}^{-3}$. The dashed lines are also contours for the less dense outer part, with the closest to the center denoting $1.8\times10^5\,\mathrm{cm}^{-3}$ and the others spaced at intervals of $0.2\times10^5\,\mathrm{cm}^{-3}$. The orange and blue dashed circles represent the Bondi and BHL radius, respectively.}
  \label{figure:snap_sub}
\end{figure*}

\begin{figure*}[htb]
  \centering
  \includegraphics[keepaspectratio, scale=0.54]{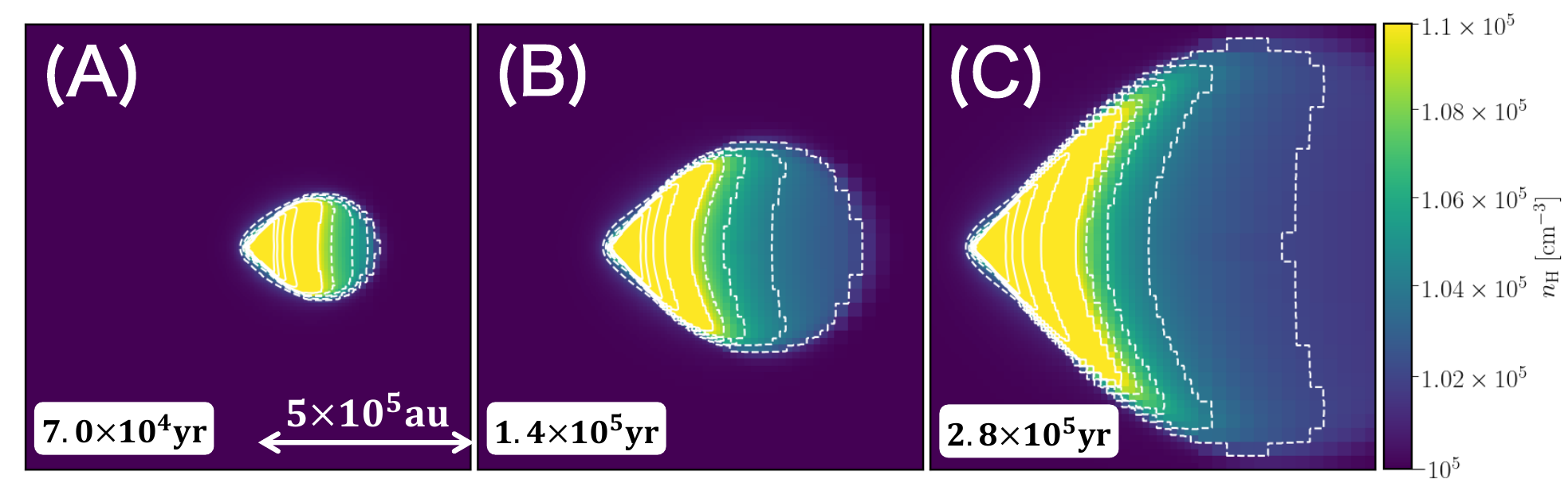}
  \caption{Same as Figure~\ref{figure:snap_sub} but for a supersonic case with $\mathcal{M}\simeq 1.49$. Panels (A), (B), and (C) represent the time sequence, corresponding to epochs at (A) $7\times10^4 \, \mathrm{yr}$, (B) $1.4\times10^5 \, \mathrm{yr}$, and (C) $2.8\times10^5 \, \mathrm{yr}$, respectively. The solid line closest to the center denotes $1.5\times10^5\,\mathrm{cm}^{-3}$, and the others are spaced at intervals of $10^4\,\mathrm{cm}^{-3}$. The dashed line closest to the center denotes $1.08\times10^5\,\mathrm{cm}^{-3}$, and the others are spaced at intervals of $2\times10^3\,\mathrm{cm}^{-3}$. In panels (B) and (C), the body is positioned to the left of the center to show a larger portion of the downstream side. }
  \label{figure:snap_sup}
\end{figure*}

\begin{figure}[htb]
  \centering
  \includegraphics[keepaspectratio, scale=0.65]{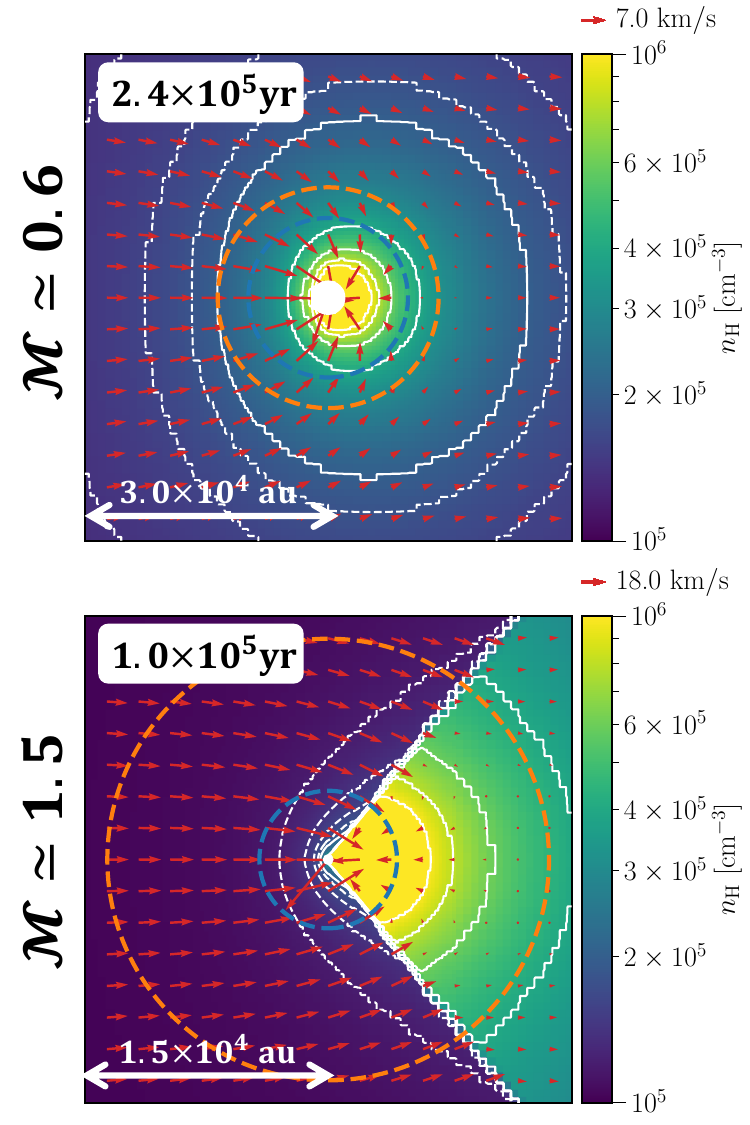}
  \caption{Enlarged views of the accretion flows for cases with $\mathcal{M}\simeq0.6$ (upper) and $\mathcal{M}\simeq1.5$ (lower). We take the snapshots at $2.4\times10^5\,\mathrm{yr}$ (the same as panel (C) in Figure~\ref{figure:snap_sub}) for the upper panel and $1.0\times10^5\,\mathrm{yr}$ for the lower panel. The red arrows indicate the gas velocity vectors, with the legend in the upper right of each panel showing their normalization.
  The solid line closest to the center denotes $1\times10^6\,\mathrm{cm}^{-3}$ and the others spaced at intervals of $2\times10^5\,\mathrm{cm}^{-3}$.  The dashed lines are also contours for the less dense outer part, with the closest to the center denoting $1.8\times10^5\,\mathrm{cm}^{-3}$ and the others spaced at intervals of $0.2\times10^5\,\mathrm{cm}^{-3}$. The orange and blue dashed circles represent the Bondi and BHL radii, respectively. In the lower panel, we present the result from the run with the sink particle size reduced by a factor of eight compared with the fiducial run (see Section~\ref{subsubsec:resolution}), to provide a better description of the inner flow structure. }
  \label{figure:snap_Bondi}
\end{figure}

Figure~\ref{figure:snap_sub} presents the temporal evolution of density perturbations induced by a body moving at a subsonic speed of $\mathcal{M}\simeq0.62$. Panels (A) to (C) describe the time evolution, illustrating the outward expansion of the region influenced by the central body. It is evident that the downstream side of the body exhibits higher density, leading to a gravitational force acting in the direction of deceleration.

Figure~\ref{figure:snap_sup} is the same as Figure~\ref{figure:snap_sub} but for a supersonic case with $\mathcal{M}\simeq1.49$. The influence region has a conical and spherical shape due to the supersonic speed of the body (see Figure~2 of \citetalias{1999ApJ...513..252O}). Similarly to the subsonic case, the downstream side of the body has a higher density, and the gravitational force acts in the direction of deceleration. 

Figure~\ref{figure:snap_Bondi} shows the same snapshots as in Figures~\ref{figure:snap_sub} (upper) and \ref{figure:snap_sup} (lower), but for enlarged views of the accretion flow near the body. This figure shows that gas is pulled by the gravitational force of the accretor and moves towards it within the BHL radius, both in subsonic and supersonic cases. Outside the BHL radius, the gas motion is still affected by the accretor's gravity, but the gas passes by the accretor without being trapped by it.

\subsubsection{Time evolution of accretion rate and friction}

\begin{figure}[htb]
  \includegraphics[keepaspectratio, scale=0.35]{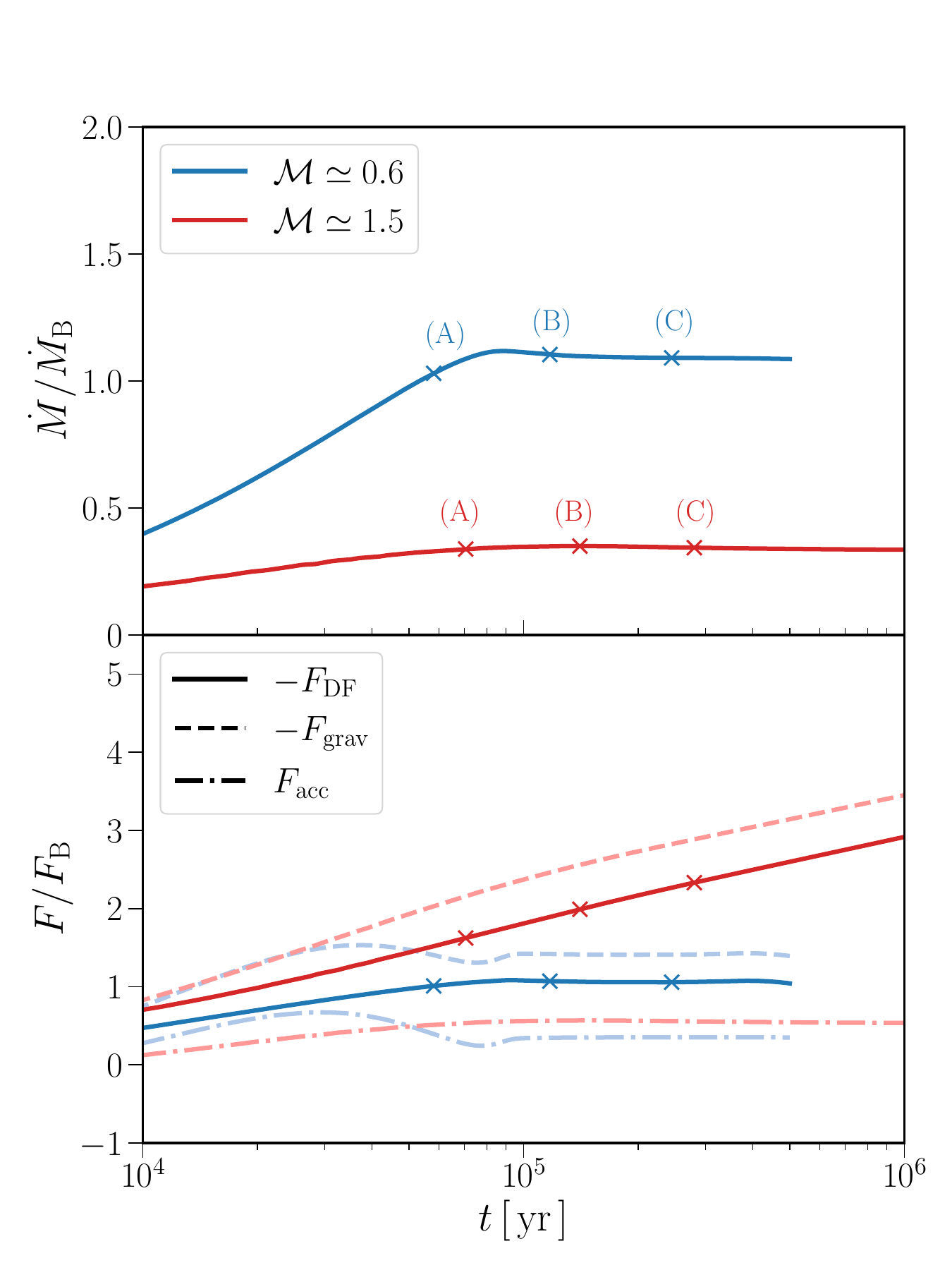}
  \caption{
The time evolution of the accretion rate (upper) and the friction force (lower). In both panels, the blue and red lines represent the same subsonic and supersonic cases as in Figures~\ref{figure:snap_sub} and \ref{figure:snap_sup}, with Mach numbers of $\mathcal{M}\simeq0.62$ and $\mathcal{M}\simeq1.49$, repsectively. The crosses on these lines correspond to the epochs (A)-(C) presented in these figures. The vertical axis in the upper panel represents the accretion rate normalized by the Bondi rate $\dot{M}_{\mathrm{B}}$ (see Equation~\ref{eq:Bondi_rate}), while the vertical axis in the lower panel represents the forces exerted on the body normalized by $\FB = 4 \pi (GM)^2 \rho_\infty / c_s^2$ (see Equation~\ref{eq:Bondi_flux}).
The solid line represents the net frictional force $-F_{\mathrm{DF}}$, which is the difference between the gravity force from density perturbations $-F_{\mathrm{grav}}$ (dashed) and the momentum flux of the accreting gas $F_{\mathrm{acc}}$ (dot-dashed line). Note that $-F_{\mathrm{DF}}$ becomes constant for $t \gtrsim 10^5$~ years in the subsonic case, and it continues to increase in the supersonic case. 
}
  \label{figure:time_dep}
\end{figure}

The top panel of Figure~\ref{figure:time_dep} displays the time evolution of the accretion rate. In both subsonic and supersonic cases, the accretion rates take nearly constant values after approximately $10^5\,\mathrm{yr}$, indicating that the accretion flows reach quasi-steady states. 

The bottom panel of Figure~\ref{figure:time_dep} describes the time evolution of the frictional force. In this panel, 
we plot the net frictional force $F_{\mathrm{DF}}$ (Equation~\ref{eq:net}), together with the gravitational force by surrounding gas $F_{\mathrm{grav}}$ (Equation~\ref{eq:DF}) and the momentum flux transferred by accreting gas $F_{\mathrm{acc}}$ (Equation~\ref{eq:momentum}). We put minus signs on $F_{\mathrm{DF}}$ and $F_{\mathrm{grav}}$, as $F_{\mathrm{DF}}$ and $F_{\mathrm{grav}}$ are negative (causing deceleration) while $F_{\mathrm{acc}}$ is positive (causing acceleration).

In the subsonic case, $F_{\mathrm{grav}}$ approaches a constant value, which is consistent with the linear perturbation theory (see Section~\ref{sec:theory}). The momentum flux $F_{\mathrm{acc}}$ also becomes constant after the accretion rate becomes constant. As a result, the net frictional force, $F_{\mathrm{DF}}=F_{\mathrm{grav}}+F_{\mathrm{acc}}$, converges to a constant value in the quasi-steady state. Hereafter, we consider this converged value as dynamical friction in the subsonic case.

In the supersonic case, $-F_{\mathrm{grav}}$ increases with time in proportion to $\log{t}$, as predicted by the linear theory (see Section~\ref{sec:theory}). On the other hand, $F_{\mathrm{acc}}$ becomes constant after the gas accretion rate becomes constant, as in the subsonic case. Consequently, $-F_{\mathrm{DF}}$ increases with time, reflecting the time dependence of $F_{\mathrm{grav}}$.

\subsubsection{Sink size dependence of friction} \label{subsubsec:resolution}

\begin{figure}[tb]
  \includegraphics[keepaspectratio, scale=0.4]{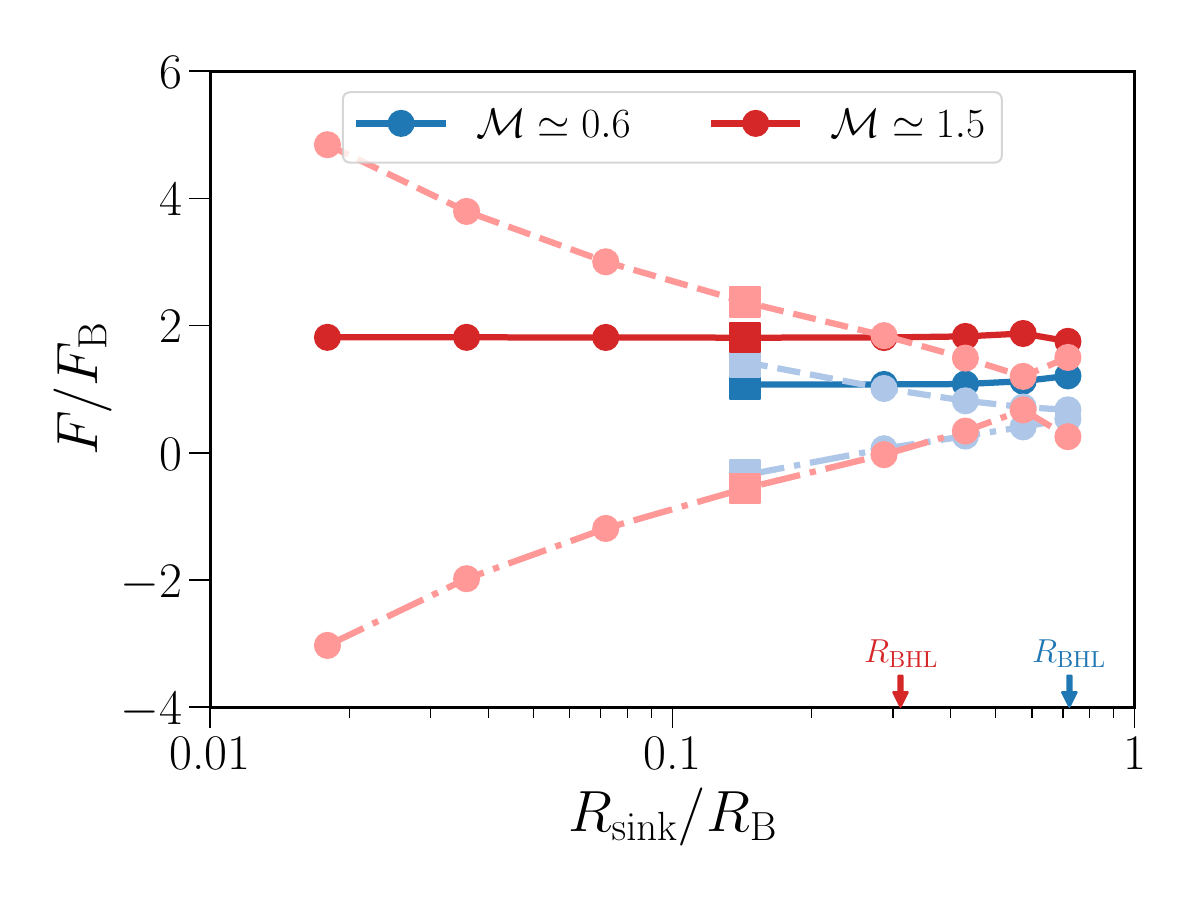}
  \caption{Dependence of the gravitational force from the gas 
  $-F_{\mathrm{grav}}$ (dashed line), the momentum flux carried by the accretion gas $F_{\mathrm{acc}}$ (dot-dashed line), and the net frictional force $-F_{\mathrm{DF}} = -(F_\mathrm{grav} + F_\mathrm{acc})$ (solid line) on the sink radius. The horizontal axis represents the sink radius normalized by the Bondi radius $R_\mathrm{B}$, and the vertical axis represents the force exerted on the body normalized by $\FB=4\pi(GM)^2\rho_{\infty}/\cs^2$ (see
    Equation~\ref{eq:Bondi_flux}).  
  The different colors represent different values of $\mathcal{M}$, where the blue and red lines correspond to $\mathcal{M}\simeq0.62$ and $\simeq 1.49$, respectively. We evaluate the gravitational and net frictional forces after they reach constant values in the case with $\mathcal{M}\simeq0.62$ and
  at $t=1.0\times10^5\,\mathrm{yr}$ in the case with $\mathcal{M}\simeq1.49$.  
  The square symbols represent our fiducial choice of the sink radius, $R_\mathrm{sink}=0.14 R_\mathrm{B}$ (Table~\ref{table:parameter}). The vertical arrows indicate the positions of the BHL radii with $\mathcal{M}\simeq 0.62$ and $1.49$. For each case of $\mathcal{M}\simeq 0.62$ and $1.49$, while $-F_\mathrm{grav} (F_{\mathrm{acc}})$ increases (decreses) monotonically with decreasing sink radius, $-F_{\mathrm{DF}}$ converges to a constant value for $R_{\mathrm{sink}} \ll R_{\mathrm{BHL}}$.}
  \label{figure:sink_fri}
\end{figure}

To assess the numerical convergence of the frictional force with increasing resolution, we examine the impact of varying the size of the sink particle on the magnitude of friction. 
Figure~\ref{figure:sink_fri} shows the magnitudes of the gravitational force and the dynamical friction as functions of the sink radius. We see that the magnitude of gravitational force increases as the sink radius decreases, whereas the magnitude of dynamical friction remains nearly constant. 

The gravitational force $F_{\mathrm{grav}}$ is affected by the size of the sink. According to Section~\ref{sec:theory}, the gas inside the BHL radius is approximately in free fall, and thus the density profile can be expressed as $\rho \propto r^{-3/2}$, where $r$ is the distance from the body. Ignoring the angular dependence of the radial distribution of the density, we can roughly estimate $| F_{\mathrm{grav}} |$ as 
\begin{align} 
|F_{\mathrm{grav}}| \sim \int \dd\mathrm{V} \frac{GM\rho}{r^2} \propto R_\mathrm{sink}^{-1/2}.
\end{align} 
This means that, as the sink size decreases, the body experiences a stronger gravitational force from the high-density gas closer to it. As a result, $F_{\mathrm{grav}}$ is resolution dependent and cannot be considered a physical force.

The cause of net frictional force being independent of the sink radius is as follows. The gas inside the BHL radius is mainly absorbed by the body, transferring momentum to it. This momentum counteracts the gravitational force from the gas that is enclosed in the BHL radius. Consequently, the sum of the gravitational force and the momentum flux does not depend on the sink radius. We consider this combined force to be the dynamical friction. We also verified that the dynamical friction converges to a constant value for other cases with $\mathcal{M} \simeq 1.29, 1.73, 2.23$, and 2.73, although this is not shown in the figure for simplicity.

\subsection{Mach number dependence of dynamical friction}

In what follows, we investigate the Mach number dependence of the dynamical friction, considering all the cases with different Mach numbers (Table~\ref{table:parameter}).

\begin{figure}[tb]
  \begin{center}
    \includegraphics[keepaspectratio, scale=0.4]{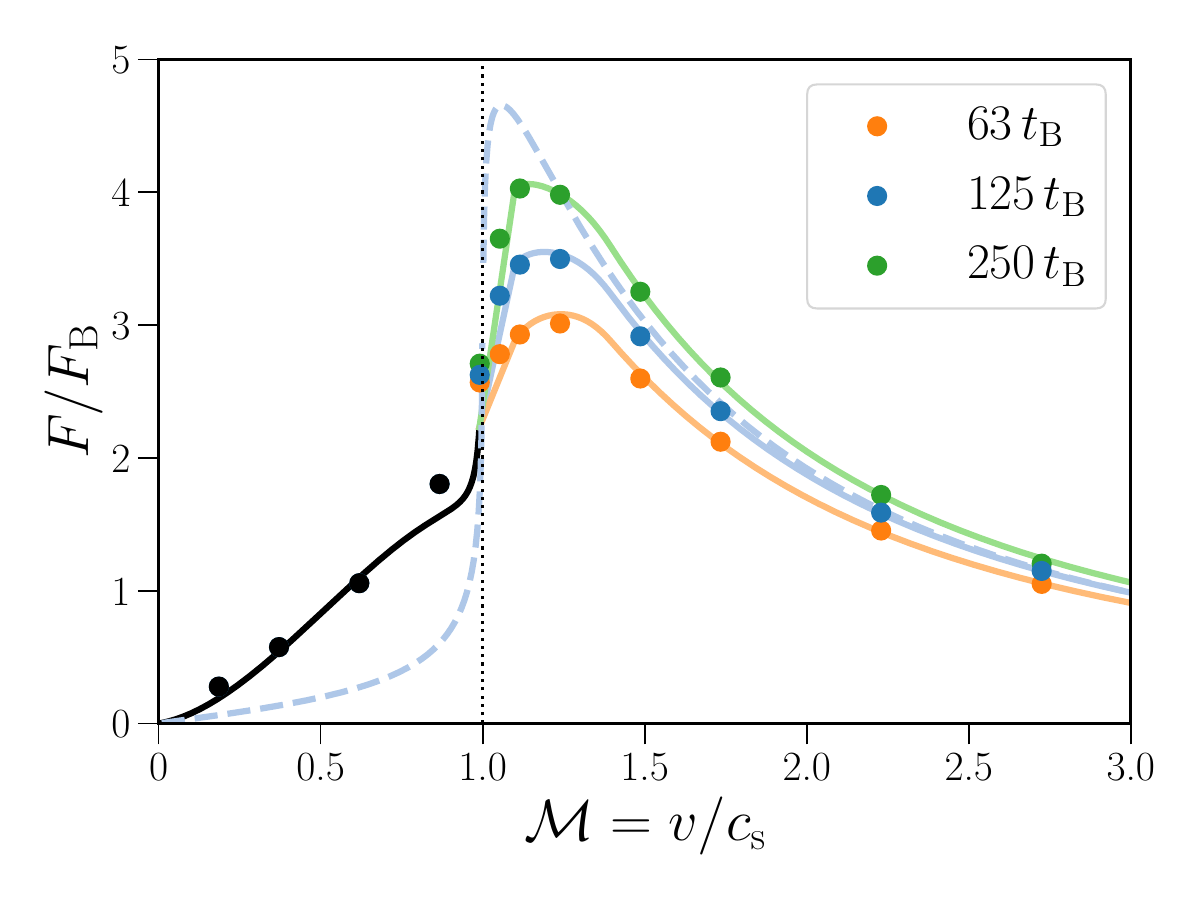}
    \caption{Velocity dependence of dynamical friction, obtained by the
    sum of gravitational force and momentum flux.  The vertical axis
    represents the magnitude of dynamical friction $|F_{\mathrm{DF}}|$
    normalized by $\FB=4\pi(GM)^2\rho_{\infty}/\cs^2$ (see
    Equation~\ref{eq:Bondi_flux}) and the horizontal axis represents the
    Mach number $\mathcal{M}$.  The different colors denote different
    periods: orange, blue, and green represent
    $t=63\,t_{\mathrm{B}}\,(5\times10^5\,\mathrm{yr})$,
    $125\,t_{\mathrm{B}}\,(1\times10^6\,\mathrm{yr})$, and
    $250\,t_{\mathrm{B}}\,(2\times10^6\,\mathrm{yr})$, respectively,
    with black corresponding to time-independent quantities.  Here,
    $t_{\mathrm{B}}=R_{\mathrm{B}}/\cs$ is the sound crossing time at
    the Bondi radius (see Equation~\ref{eq:t_B}).  In subsonic cases, except for $\mathcal{M} \simeq
    0.99$, we plot the dynamical friction at
    $25\,t_{\mathrm{B}}\,(2\times10^5\,\mathrm{yr})$ for
    $\mathcal{M}\simeq0.19, 0.37, 0.62$ and at
    $63\,t_{\mathrm{B}}\,(5\times10^5\,\mathrm{yr})$ for
    $\mathcal{M}\simeq0.87$, when the dynamical friction has already reached
    a constant value.  We overplot our fitting formula given by
    Equation~\eqref{eq:fit_form} for $t=63$, $125$, and $250\,t_{\mathrm{B}}$
    (solid lines) and the linear formula from \citetalias{1999ApJ...513..252O} given by
    Equation~\eqref{eq:DF_lin} with $r_{\mathrm{min}}=\RBHL/2$ for
    $125\,t_{\mathrm{B}}$ (dashed line).}\label{figure:Mach_Fri}
  \end{center}
\end{figure}

Figure~\ref{figure:Mach_Fri} illustrates the variation of the net frictional force (dynamical friction) with different Mach numbers.
In the range of $\mathcal{M} > 1$, different colors of the data points represent different epochs of $63\,t_{\mathrm{B}}\,(5 \times 10^5\,\mathrm{yr, orange})$, 
$125\,t_{\mathrm{B}}\,(1 \times 10^6\,\mathrm{yr, blue})$, 
and $250\,t_{\mathrm{B}}\,(2 \times 10^6\,\mathrm{yr, green})$, where
\begin{align}
    t_{\mathrm{B}} &= R_{\mathrm{B}}/\cs \nonumber\\
    &= 8.0\times10^3\,\mathrm{yr}\,\bigg(\frac{M}{10^3\,M_{\odot}}\bigg)
    \bigg(\frac{T_0}{10^4\,\mathrm{K}}\bigg)^{-3/2}
    \bigg(\frac{\mu}{1.26}\bigg)^{3/2}.
    \label{eq:t_B}
\end{align}
For the subsonic cases, except for $\mathcal{M} \simeq 0.99$, the black dots represent the values in the quasi-steady state. 
We also compare our simulations with the linear perturbation theory \citepalias{1999ApJ...513..252O} with the time fixed to $t=125\,t_{\mathrm{B}}$. In Figure~\ref{figure:Mach_Fri}, the blue dashed line represents the linear perturbation theory, for which we choose  $r_{\mathrm{min}} = \RBHL/2$ to give an overall fit to the simulation data (see Equation~\ref{eq:DF_lin}-\ref{eq:Bondi_flux}). Figure~\ref{figure:Mach_Fri} shows that on the subsonic side, the simulation results tend to be higher than those of linear perturbation theory. On the supersonic side, the simulation results are smaller than those predicted by the linear perturbation theory. However, the difference decreases as the Mach number increases. 

We provide a formula that fits our simulation results much better than the linear perturbation theory with $r_{\mathrm{min}}=R_{\mathrm{BHL}}/2$ but is based on it,
\begin{align}
    F = F_{\mathrm{B}}\mathcal{I}(\mathcal{M}),  
    \label{eq:fit_form}
\end{align}
where $\mathcal{I}(\mathcal{M})$ takes distinct forms in different regimes of $\mathcal{M}$. First, in the subsonic regime of $0<\mathcal{M}<0.99$, 
\begin{align}
    \label{eq:subsonic}
\mathcal{I}(\mathcal{M})
= \frac{\lambda(\mathcal{M})}{\mathcal{M}^2}
        \Biggl[ \frac{1}{2}\ln(\frac{1+\mathcal{M}}{1-\mathcal{M}})-\mathcal{M} \Biggr],
\end{align}
where
\begin{align}
        \lambda(\mathcal{M}) = -12\,\mathcal{M}^2 + 12\,\mathcal{M} + \frac{6}{5}\,.
        \label{eq:lambda_sub}
\end{align}
The functional shape of Equation\eqref{eq:subsonic} is the same as that of linear theory Equation~(\ref{eq:Ostriker}) for $\mathcal{M} < 1$ except for the factor of $\lambda(\mathcal{M})$. We introduce this factor to enhance the frictional force with respect to the linear values. Our simulation results show the largest deviation from the linear theory around $\mathcal{M}=0.5$, where $\lambda(\mathcal{M})$ takes its maximum value.
Next, in the supersonic regime of $\mathcal{M}>1.1$, 
    \begin{equation}
    \label{eq:supersonic}
        \begin{split}
            \mathcal{I}(\mathcal{M}) &= \frac{1}{\mathcal{M}^2}
            \Biggl[ \frac{1}{2} \ln(1-\frac{1}{\mathcal{M}^2})\\
            &\quad + \ln(\frac{vt}{\zeta(\mathcal{M})\RBHL}) \Biggr],
        \end{split}
    \end{equation}
where
    \begin{equation}
    \label{eq:zeta}
        \begin{split}
            \zeta(\mathcal{M}) &= 
            \begin{cases}
                \displaystyle
                14\,\mathcal{M}^2 - \frac{196}{5}\,\mathcal{M} 
                + \frac{281}{10}, & 1.1 < \mathcal{M} \leq 1.4, \vspace{0.2cm}\\
                \displaystyle
                -\frac{1}{10}\mathcal{M} + \frac{4}{5}, & 1.4 < \mathcal{M} \leq 3.0, \vspace{0.2cm}\\
                \displaystyle
                \frac{1}{2}, & \mathcal{M} > 3.0.
            \end{cases}
        \end{split}
    \end{equation}
This formula also has the same form as that of Equation~(\ref{eq:Ostriker}) for $\mathcal{M}>1$ but with the parameter $r_{\mathrm{min}}$ replaced by 
the BHL radius $\RBHL$ multiplied by $\zeta(\mathcal{M}) \sim \mathcal{O}(1)$.
The factor $\zeta(\mathcal{M})$ decreases with $\mathcal{M}$
for $1.1<\mathcal{M}\leq3.0$ and takes a constant value of $1/2$ for the larger Mach numbers.
Finally, in the intermediate regime for $0.99\leq\mathcal{M}\leq1.1$, 
    \begin{align}
    \label{eq:middle}
        \mathcal{I}(\mathcal{M}) = a(t)\mathcal{M}+b(t),
    \end{align}
where 
    \begin{equation}
        \begin{split}
            \begin{cases}
                \displaystyle
                a(t) &= \displaystyle\frac{10000}{1331}\bigg(A(t)-\frac{1099}{675}B\bigg), 
                \vspace{0.2cm}\\
                \displaystyle
                b(t) &= \displaystyle-\frac{900}{121}\bigg(A(t)-\frac{2198}{1215}B\bigg),
            \end{cases}
        \end{split}
    \end{equation}
and
    \begin{equation}
    \label{eq:coefficient}
        \begin{split}
            \begin{cases}
                \displaystyle
                A(t) &= \displaystyle\ln{\bigg(\frac{\cs^3t}{GM}\bigg)} + \ln{\bigg(\frac{221\sqrt{7}}{640\sqrt{3}}\bigg)}, \vspace{0.2cm}\\
                \displaystyle
                B &= \displaystyle\frac{1}{2}\ln{199} - \frac{99}{100}.
            \end{cases}
        \end{split}
    \end{equation}
The linear function of $\mathcal{M}$ in the intermediate regime (Equation~\ref{eq:middle}) connects
the function in the subsonic (Equation~\ref{eq:subsonic}) and supersonic (Equation~\ref{eq:supersonic}) regimes, making the fitting formula provided by Equation~(\ref{eq:fit_form}) a continuous function of Mach number across the entire Mach number range.

The above results suggest that, in the supersonic regime, we can fix the undetermined parameter in the linear theory $r_{\mathrm{min}}$, by applying the linear formula to reproduce the simulation results. Equations~(\ref{eq:supersonic}) and (\ref{eq:zeta}) indicate that $r_{\mathrm{min}}$ determined in such a way should be similar to the BHL radius, a comparison that will be given in Section~\ref{subsubsec:supersonic}. Here, we consider how our results are compared with \citet{1999ApJ...522L..35S} and \citet{2009ApJ...703.1278K}, who also conducted numerical simulations to study the dynamical friction exerted on the linearly moving body. Note that both of these authors assumed a gravitational softening radius $R_{\mathrm{soft}}$ significantly larger than the Bondi radius, neglecting the effect of mass accretion, unlike ours. As a result, they confirmed that their results are well described by the linear theory assuming $r_{\mathrm{min}}$ sufficiently larger than the Bondi radius. 
For example, \citet{2009ApJ...703.1278K} gave $r_{\mathrm{min}}=0.35 \mathcal{M}^{0.6}R_{\mathrm{soft}}$, which corresponds to $35~R_{\mathrm{B}} \leq r_{\mathrm{min}} \leq 68~R_{\mathrm{B}}$ within the range of $1.0 \leq \mathcal{M} \leq 3.0$.
This is in stark contrast to our results, where the gravitational softening radius is much smaller than the Bondi radius, and the influence of mass accretion is of central interest. 

So far, we have presented the overall Mach number dependence of dynamical friction. In the subsequent sections, to understand the obtained Mach number dependence, we examine how the dynamical friction is caused in our simulations comparing our results with the linear perturbation theory. We describe the subsonic case in Section~\ref{subsubsec:subsonic} and the supersonic case in Section~\ref{subsubsec:supersonic}.

\subsubsection{Subsonic cases}
\label{subsubsec:subsonic}

\begin{figure}[tb]
  \centering
  \includegraphics[keepaspectratio, scale=0.6]{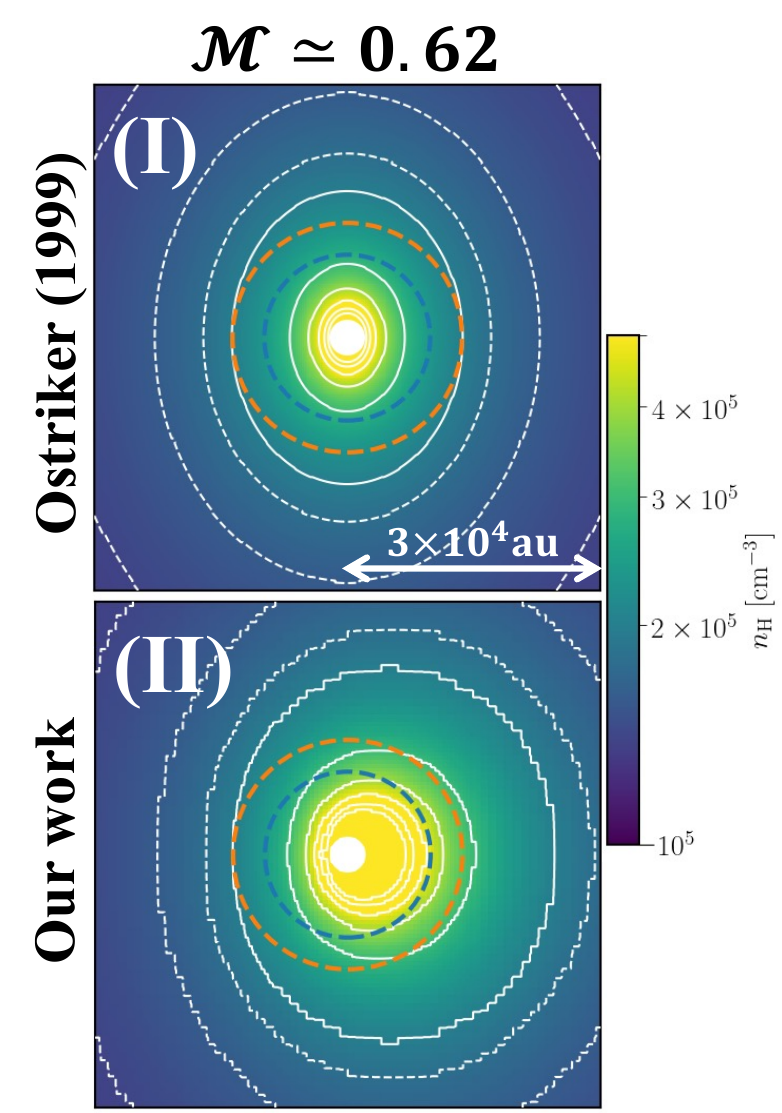}
  \caption{Comparison of density distributions in the linear theory (panel I) and simulation (panel II)
   for the case with $\mathcal{M}\simeq0.62$ at $t=2.4\times10^5\,\mathrm{yr}$ (same as panel (C) in Figure\ref{figure:snap_sub}). Panel (I) shows the density predicted by the linear perturbation theory \citepalias{1999ApJ...513..252O}, while Panel (II) shows that from our simulation. The white filled circle at the center illustrates the sink particle and white lines indicate the isodensity contours between $2\times10^5\,\mathrm{cm}^{-3}$ and $7\times10^5\,\mathrm{cm}^{-3}$ with intervals of $10^5\,\mathrm{cm}^{-3}$ (solid) and between $1.8\times10^5\,\mathrm{cm}^{-3}$ and $1.4\times10^5\,\mathrm{cm}^{-3}$ with intervals of $2\times10^4\,\mathrm{cm}^{-3}$ (dashed). The plotted region has a side length of  $6\times10^4\,\mathrm{au}$, and Panel (II) corresponds to a three-times zoom-in of panel (C) in Figure\ref{figure:snap_sub}. The orange and blue dashed lines represent the positions of the Bondi and BHL radii, respectively.}
  \label{figure:den_sub}
\end{figure}

First, we consider why the value of dynamical friction in the subsonic case is larger than that obtained from the linear perturbation theory (see Figure~\ref{figure:Mach_Fri}). Figure~\ref{figure:den_sub} compares the gas number density $n_\mathrm{H}$ near the center in the case of $\mathcal{M}\simeq0.62$ predicted by the linear perturbation theory \citepalias[panel (I);][]{1999ApJ...513..252O} and obtained from our simulation (panel (II); at $t=2.4\times10^5\,\mathrm{yr}$, same as Figure~\ref{figure:snap_sub} (C)).

\begin{figure}[tb]
  \centering
  \includegraphics[keepaspectratio, scale=0.4]{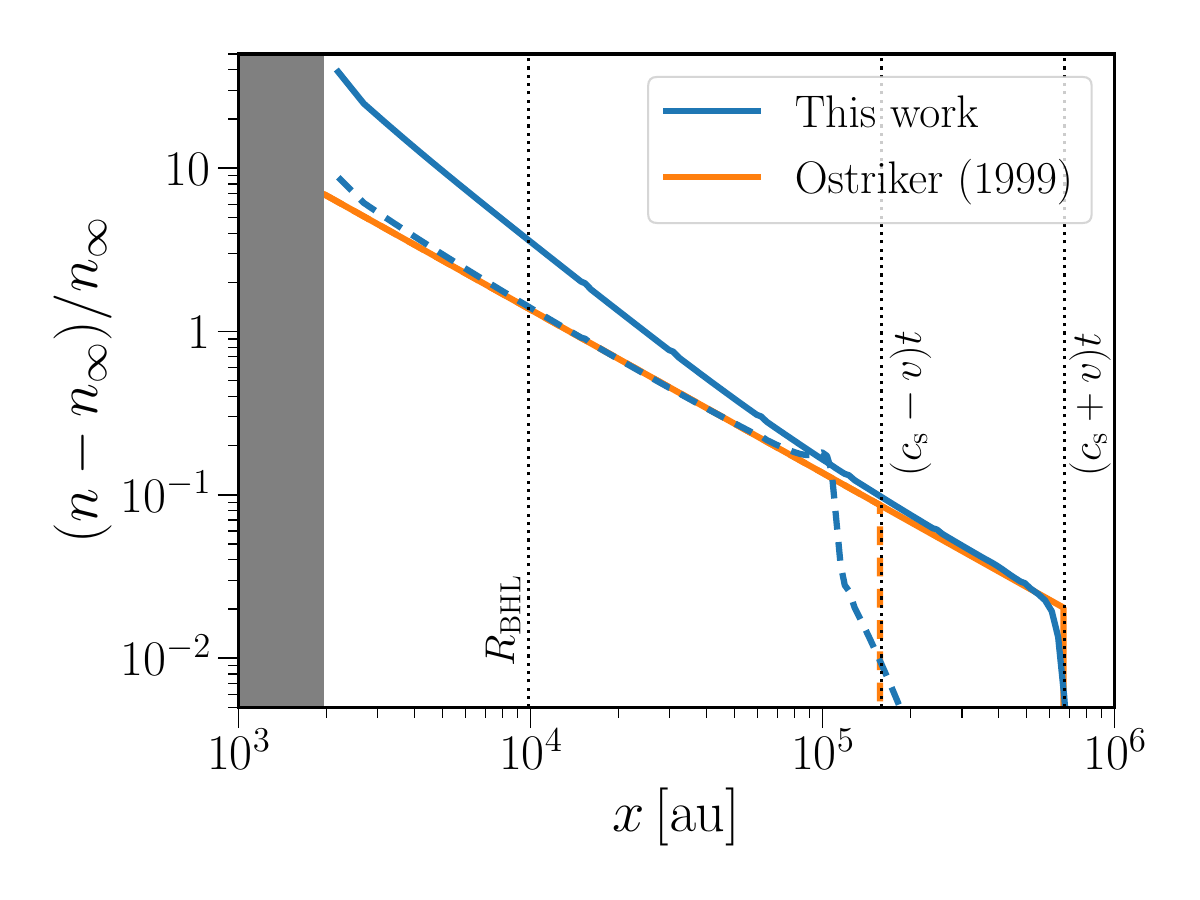}
  \caption{Density profiles along the $x$-axis for the case with $\mathcal{M}\simeq0.62$
   at the time of Figure\ref{figure:snap_sub} (C). The horizontal axis represents the distance $x$ from the central body, while the vertical axis represents the normalized difference between the density at $x$ and its ambient value, i.e., $(n-n_{\infty})/n_{\infty}$. We show the density obtained from our simulations with blue lines and that obtained from the linear perturbation theory \citepalias{1999ApJ...513..252O} with orange lines. Solid and dashed lines represent values for $x>0$ and $x<0$, respectively. Vertical dotted lines indicate, from left to right, the BHL radius $R_{\mathrm{BHL}}$, $(\cs-v)t$, and $(\cs+v)t$. The region shaded in gray represents the location of a sink particle. Note that in the linear perturbation theory, the density distribution is symmetrical for $|x|<(\cs-v)t$, leading to identical density profiles for $x>0$ and $x<0$.}
  \label{figure:den_dis_sub}
\end{figure}

In Figure~\ref{figure:den_sub} (I), the isodensity contours in the linear perturbation theory
form symmetric elliptical shapes centered at the body. 
In Figure~\ref{figure:den_dis_sub}, we plot the gas number density profiles along the $x$-axis in both forward and backward directions with respect to the body's motion. The profiles show that the symmetric structure observed in Figure~\ref{figure:den_sub} (I) extends up to $x=(\cs-v)t$ on the $x$-axis. Consequently, in the linear theory, the gravitational force exerted by the gas for $x<(\cs-v)t$ cancels out due to backward-forward symmetry and friction is caused by the asymmetry in the gas density for $x>(\cs-v)t$. 

In contrast, in Figure~\ref{figure:den_sub} (II),
the centers of the isodensity contours in our simulation are located downstream side of the body. This positional shift occurs because accretion of gas onto a body moving at a constant velocity takes place primarily on the downstream side, leading to an accumulation of gas in the downstream direction. Thus, the density on the downstream side becomes higher than on the upstream side. In Figure~\ref{figure:den_dis_sub}, the density profiles along the $x$-axis exhibit asymmetric even in the region with $x<(\cs-v)t$, where the gas distribution is symmetric in the linear theory. Hence, in our simulation, the gravitational force from the region of $x<(\cs-v)t$ does not cancel out, unlike in the linear theory as mentioned above. This explains why the dynamical friction is greater than that predicted by the linear theory. Recall that the gas within $x<R_{\mathrm{BHL}}$ does not contribute to dynamical friction because mass accretion compensates for the gravitational force. Only the gas in the range $R_{\mathrm{BHL}} < x < (\cs-v)t$ provides an additional contribution to dynamical friction, compared to the linear theory.

\subsubsection{Supersonic cases}
\label{subsubsec:supersonic}

\begin{figure}[tb]
  \centering
  \includegraphics[keepaspectratio, scale=0.4]{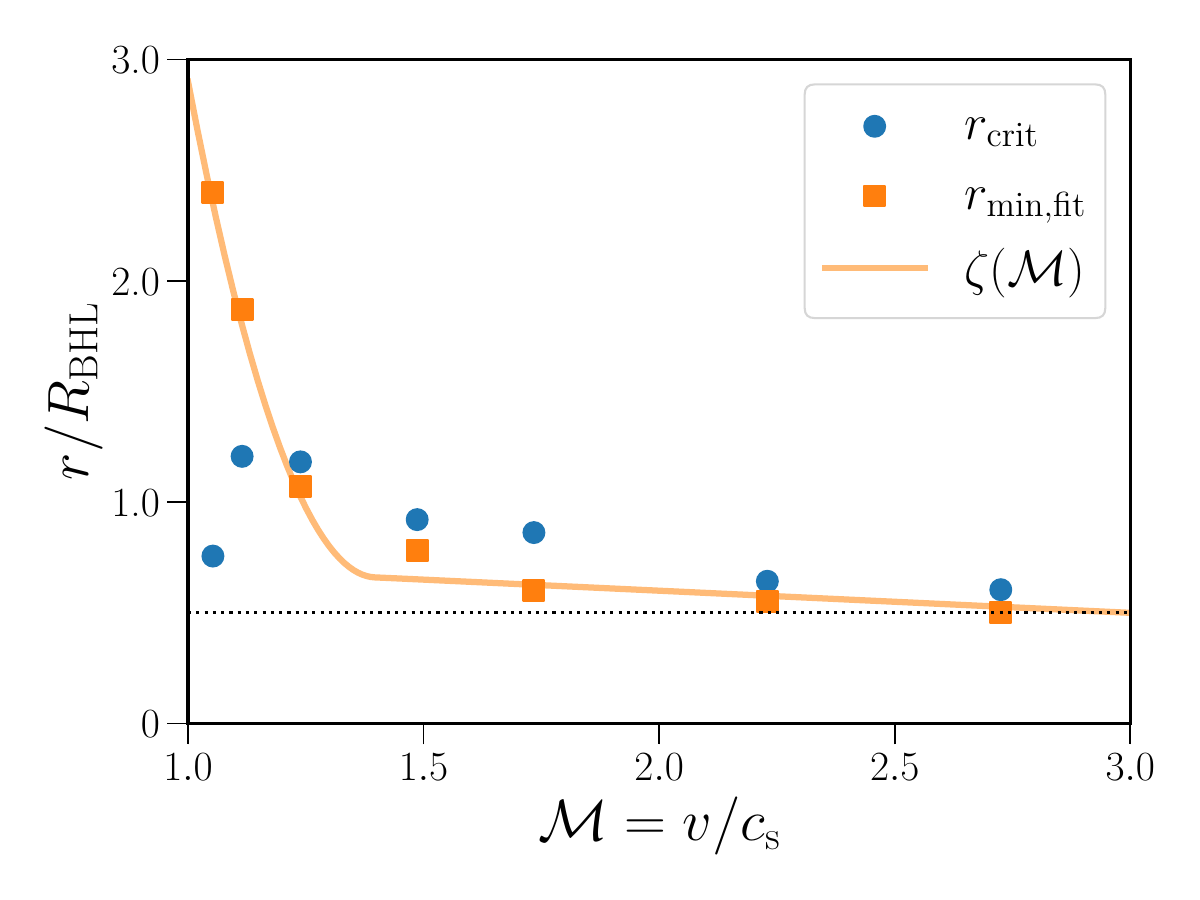}
  \caption{The Mach number dependence of characteristic length scales in our simulations for the supersonic cases. The blue circles represent the critical radius $r_{\mathrm{crit}}$, where the momentum flux carried by the accreting gas cancels out the gravitational force exerted by the gas within a radial distance $r$, $F_{\mathrm{DF}}(<r) = 0$ (Equation~\ref{eq:r_crit}). The orange squares represent the minimum radius $r_{\mathrm{min, fit}}$, which is determined so that the linear formula (Equation~\ref{eq:Ostriker}) fits the dynamical friction measured in our simulations, $F_\mathrm{DF}$ (Equation~\ref{eq:net}). These radii divided by $R_{\mathrm{BHL}}$ are plotted as functions of the Mach number $\mathcal{M}$ in the figure. The orange curve illustrates $\zeta(\mathcal{M})=r_{\mathrm{min}}/R_{\mathrm{BHL}}$ given by Equation~\eqref{eq:zeta}, which is a fitting function for the Mach number dependence of $r_{\mathrm{min,fit}}$. The horizontal dotted line represents $r/R_{\mathrm{BHL}} = 1/2$. }
  \label{figure:sign_friction}
\end{figure}

\begin{figure*}[tb]
    \centering
    \includegraphics[keepaspectratio, scale=0.6]{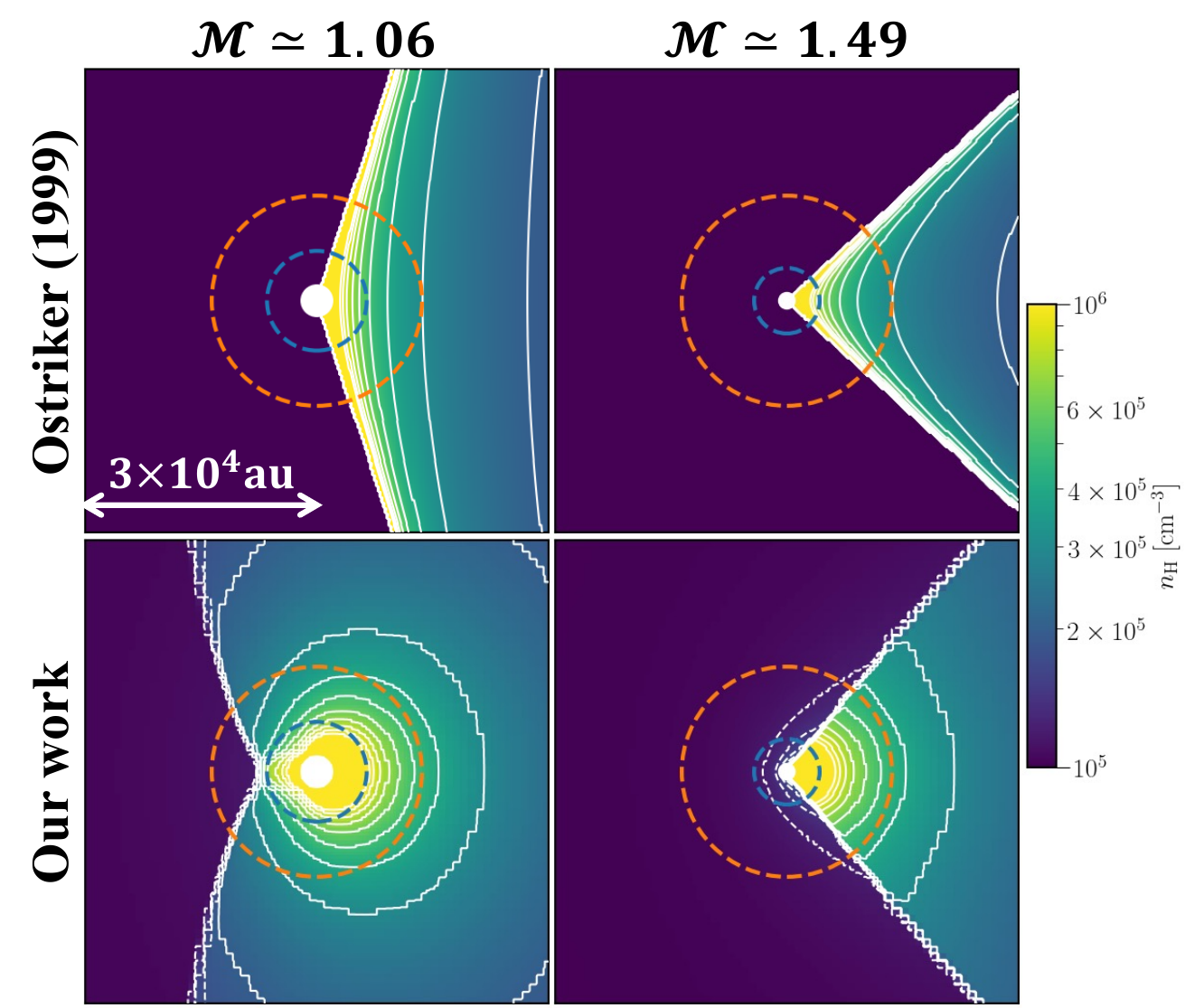}
    \caption{The same as shown in Figure~\ref{figure:den_sub}, but for the supersonic cases where $\mathcal{M}\simeq1.06$ (left) and $\mathcal{M}\simeq1.49$ (right) in an epoch of $t \simeq 1.5\times10^6 \, \mathrm{yr}$. Note that for the case with $\mathcal{M}\simeq1.49$, the result obtained from the run with a sink radius that is half the fiducial size is presented. This run provides a more detailed view of the density structure near the central body, although the value of dynamical friction does not change significantly when the sink radius is halved, as described in Section~\ref{subsubsec:resolution}.}
  \label{figure:den_sup}
\end{figure*}

\begin{figure*}[tb]
    \centering
    \includegraphics[keepaspectratio, scale=0.5]{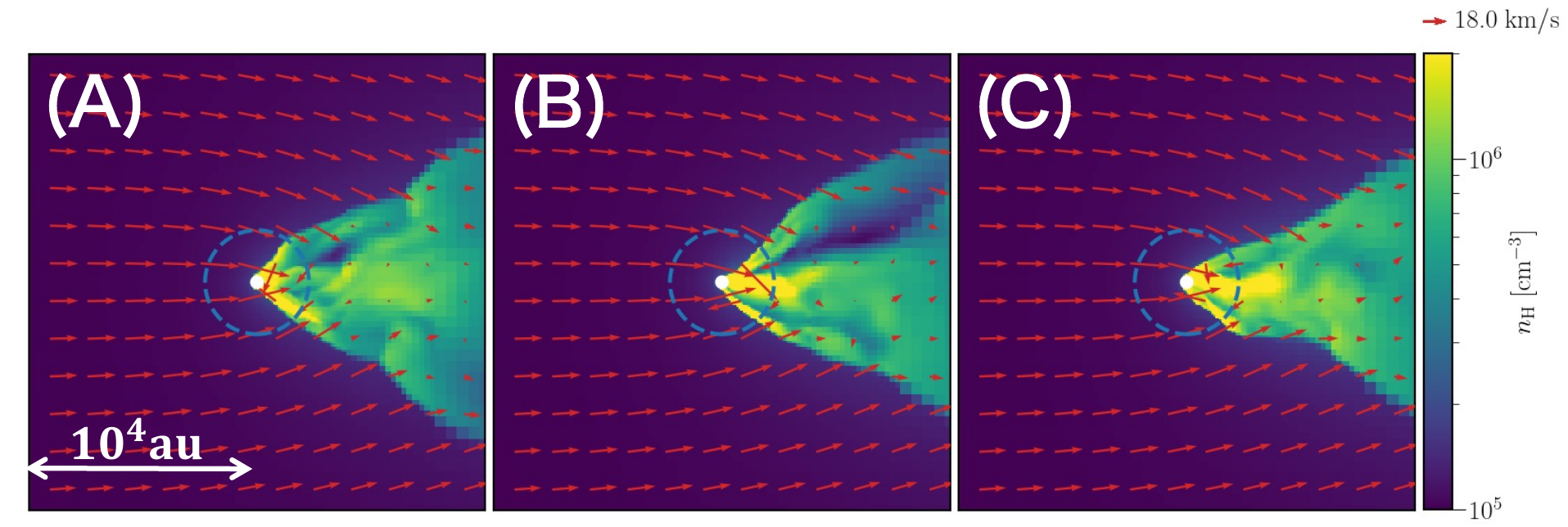}
    \caption{Non-steady flow patterns observed in the case with $\mathcal{M}\simeq2.23$. 
    We show the time evolution of the density distribution in panels (A)-(C), corresponding to $7.0\times10^4\,\mathrm{yr}$, $9.0\times10^4,\mathrm{yr}$, and $1.1\times10^5\,\mathrm{yr}$, respectively.
    Here, we present the results from the run with $R_{\mathrm{sink}}=(1/4)R_{\mathrm{sink,fiducial}}$. The blue dashed circles denote the BHL radius.}
  \label{figure:snap_instability}
\end{figure*}

We have used the identical functional form from the linear theory in our fitting formula for the supersonic cases (Equation~\ref{eq:supersonic}), albeit with the substitution of $r_{\mathrm{min}}$ with $\zeta(\mathcal{M}) R_{\mathrm{BHL}}$. We note that, in the linear theory, $r_{\mathrm{min}}$ represents the minimum radius of the gas that contributes to the dynamical friction. Using the linear theory formula to fit our nonlinear simulation results, the factor $\zeta(\mathcal{M}) R_{\mathrm{BHL}}$ accounts for additional effects beyond what is considered in the linear theory.
Here we extract such additional effects by considering the Mach number dependence of $\zeta(\mathcal{M})$.

To do so, we introduce the critical radius $r_{\mathrm{crit}}$, the radius only defined with a given simulation snapshot, apart from the linear theory. 
This is the minimum radius of the gas that contributes to the dynamical friction. 
In other words, the gravitational force from the gas enclosed by $r_{\mathrm{crit}}$ cancels out with the momentum flux carried by the accreting gas. We calculate $r_{\mathrm{crit}}$ from a given simulation snapshot as follows. We compute the gravitational force in the direction of the body's motion exerted by the gas within a radial distance $r$
\begin{align}
    F_{\mathrm{grav}}(<r) := \Bigg[\int_{V(r)} \dd^3\bm{r'}\frac{\rho(\bm{r'})GM}{r'^3}\frac{\bm{r'}}{r'}\Bigg] \cdot\frac{\bm{v}}{v},
\end{align}
where $V(r)$ represents the spherical volume of the radius $r$. We consider the dynamical friction only from the gas within the volume $V(r)$ 
\begin{align}
    F_{\mathrm{DF}}(<r) := F_{\mathrm{grav}}(<r) + F_{\mathrm{acc}},
\end{align}
and obtain $r_{\mathrm{crit}}$ as the radius at which 
\begin{align}
    F_{\mathrm{DF}}(<r_{\mathrm{crit}}) = 0.
     \label{eq:r_crit}
\end{align}

Figure~\ref{figure:sign_friction} shows how the radius $r_{\mathrm{crit}}$ varies as a function of the Mach number $\mathcal{M}$. We see that $r_{\mathrm{crit}}$ is always comparable to $R_{\mathrm{BHL}}$ for the entire range examined of $1 < \mathcal{M} < 3$. 
The gas within the BHL radius $R_{\mathrm{BHL}}$ accretes in free fall and returns its momentum to the body, which does not contribute to the friction. This coincides with the definition of $r_{\mathrm{crit}}$, which explains why $r_{\mathrm{crit}} \sim R_{\mathrm{BHL}}$.

By comparing $\zeta(\mathcal{M})$ and $r_\mathrm{crit}/R_{\mathrm{BHL}}$, we can analyze the differences between our simulation results and the predictions of linear theory, which correspond to the nonlinear effects. 
If $\zeta(\mathcal{M})$ is very close to $r_\mathrm{crit}/R_{\mathrm{BHL}}$, it means that the density distribution outside $r_{\mathrm{crit}}$ in our simulation is similar to that predicted by the linear theory. In this case, the nonlinear effects are not significant. This is illustrated in Figure~\ref{figure:sign_friction}, where $\zeta(\mathcal{M})$ does not completely coincide with $r_{\mathrm{crit}}/R_{\mathrm{BHL}}$ but the deviation is not significant for $1<\mathcal{M}<3$. 
The relations $r_{\mathrm{crit}} \sim R_{\mathrm{BHL}}$ and $\zeta(\mathcal{M}) \sim r_{\mathrm{crit}}/R_\mathrm{BHL}$ explain why $\zeta(\mathcal{M}) \sim \mathcal{O}(1)$. 
In particular, for $\mathcal{M} \gtrsim 1.3$, $\zeta(\mathcal{M})$ is very close to $r_{\mathrm{crit}}/R_{\mathrm{BHL}}$. For these cases, both $\zeta(\mathcal{M})$ and $r_{\mathrm{crit}}/R_{\mathrm{BHL}}$ are only slightly higher than $1/2$. This explains why the dynamical friction obtained in our simulations is slightly lower than the prediction of linear theory when $r_{\mathrm{min}}=R_{\mathrm{BHL}}/2$ (see the blue solid and dashed lines in Figure~\ref{figure:Mach_Fri}). 

Figure~\ref{figure:sign_friction} illustrates significant discrepancies between $\zeta(\mathcal{M})$ and $r_\mathrm{crit}/R_{\mathrm{BHL}}$ for $\mathcal{M} \lesssim 1.3$, indicating that the nonlinear effect becomes more prominent as the Mach number decreases below $\mathcal{M} \sim 1.3$. This behavior of $\zeta(\mathcal{M})$ can be explained by comparing the density distributions from our simulations and the linear theory. The left-hand side of Figure~\ref{figure:den_sup} presents the density distributions obtained from the linear perturbation theory (upper) and our simulation (lower) at $\mathcal{M}=1.06$. In the linear perturbation theory, the density distribution takes the form of a conical plus spherical shape \citepalias[see Figure~2 of][]{1999ApJ...513..252O}. In contrast, the density distribution in the simulation exhibits a different shape, with the surface of density discontinuity located on the upstream side of the body. In this case, the presence of high-density regions on the upstream side results in a smaller density difference between the upstream and downstream sides. Consequently, the dynamical friction is reduced compared to the linear theory, leading to a larger value of $\zeta(\mathcal{M})$ compared to $r_{\mathrm{crit}}/R_\mathrm{BHL}$ (Figure~\ref{figure:sign_friction}). The flow structure at $\mathcal{M}=1.06$ resembles that of the subsonic case (see Figure~\ref{figure:snap_sub}). We propose that the $\mathcal{M}=1.06$ case lies in the transitional regime between the subsonic and supersonic cases.

Figure~\ref{figure:den_sup} on the right-hand side shows the case of $\mathcal{M}=1.49$. Although the shapes of the isodensity contours differ between our simulation and the linear theory, the discrepancy in the angle of the density discontinuity surface is smaller compared to the case of $\mathcal{M} \simeq 1.06$. Consequently, the friction resulting from the gas outside $r_{\mathrm{crit}}$ is comparable in both the linear theory and our simulation, leading to $\zeta(\mathcal{M}) \sim r_{\mathrm{crit}}/R_\mathrm{BHL}$ (Figure~\ref{figure:sign_friction}).

We observe that the flow structure for $\mathcal{M}\geq2$ exhibits more temporal variability compared to other cases due to the instability reported in the literature \citep[e.g.,][]{1987MNRAS.226..785M,1988ApJ...335..862F}. Figure~\ref{figure:snap_instability} shows the temporal changes in the gas distribution for the case of $\mathcal{M}=2.23$. These temporal variations result in fluctuations in $r_{\mathrm{crit}}$. However, after $t\geq1\times10^5\,\mathrm{yr}$, $r_{\mathrm{crit}}$ stabilizes at a constant value, with fluctuations of approximately $10\%$. To obtain the values of $r_{\mathrm{crit}}$ shown in Figure~\ref{figure:sign_friction}, we average the results over six different time intervals, spaced roughly every $1\times10^4\,\mathrm{yr}$ within the range of $(1.0 - 1.5)\times10^5\,\mathrm{yr}$.

\section{Discussions} \label{sec:discussion}

\subsection{Application of friction formula}
\label{sec:application}

Our fitting formula for the dynamical friction, Equation~\eqref{eq:fit_form}, can be applied to a variety of problems. It is an updated version of the formula proposed by \citetalias{1999ApJ...513..252O}, which has been widely used in the literature. One example of its application is in modeling the motion of a BH within a galaxy, as observed in cosmological simulations with limited spatial resolution. For instance, \citet{2013MNRAS.433.3297D} employ the formula 
\begin{align}
F_{\mathrm{DF}} = \frac{4\pi\alpha\bar{\rho}(GM)^2}{\bar{c}_{\mathrm{s}}}\mathcal{I}_{\mathrm{lin}}(\mathcal{M}),
\end{align} 
where $\alpha$ $(\geq 1)$ represents a boost factor that accounts for the unresolved dense gas in the vicinity of the BH, $\bar{\rho}$ and $\bar{c}_{\mathrm{s}}$ denote the mean density and mean sound speed of the resolved gas, and $\mathcal{I}_{\mathrm{lin}}(\mathcal{M})$ is the same function as defined in Equation~\eqref{eq:Ostriker}. Our formula can also be applied to the above equation by substituting $\mathcal{I}_{\mathrm{lin}}(\mathcal{M})$ with $\mathcal{I}(\mathcal{M})$ (Equation~\ref{eq:fit_form}). One of its primary effects is to systematically enhance the friction force within the subsonic range, compared to the linear theory presented in Equation~\eqref{eq:Ostriker}.

In the supersonic range, the function $\mathcal{I}(\mathcal{M})$ is time-dependent. \citet{2013MNRAS.433.3297D} set $\ln{\Lambda}=\ln(r_\mathrm{max}/r_\mathrm{min}) = 3.2$ based on the simulation results given by \citet{2013MNRAS.429.3114C}. Since our simulations have shown that $r_{\mathrm{min}}$ should be taken as $\zeta(\mathcal{M}) R_\mathrm{BHL}$,\footnote{
Some studies adopt $r_{\mathrm{min}} \sim R_\mathrm{BHL}$ in their formulas. For example, \citet{2022MNRAS.510..531C} employ $r_{\mathrm{min}}=GM/v^2$, which corresponds to the half of HL radius (see Equation~\ref{eq:HL}). }
one can determine $r_{\mathrm{max}}$ by some criteria. A simple choice of $r_{\mathrm{max}}$ is to take the typical size of the system, such as that of a galactic disc. When considering a body in linear motion, as in our simulations, one can substitute $r_{\mathrm{max}}=\mathcal{M}c_{\mathrm{s}}t$. For a body in circular motions, such as in a binary system, $r_{\mathrm{max}}$ can be approximated as the orbital radius \citep[e.g.][]{2007ApJ...665..432K}. However, since \citet{2007ApJ...665..432K} only focus on the linear regime, the dynamical friction of accreting bodies in circular motions still needs to be addressed in future work.

\subsection{Comparison with previous work}

\subsubsection{Accreting body in linear motion}

\begin{figure}[tb]
    \centering
    \includegraphics[keepaspectratio, scale=0.4]{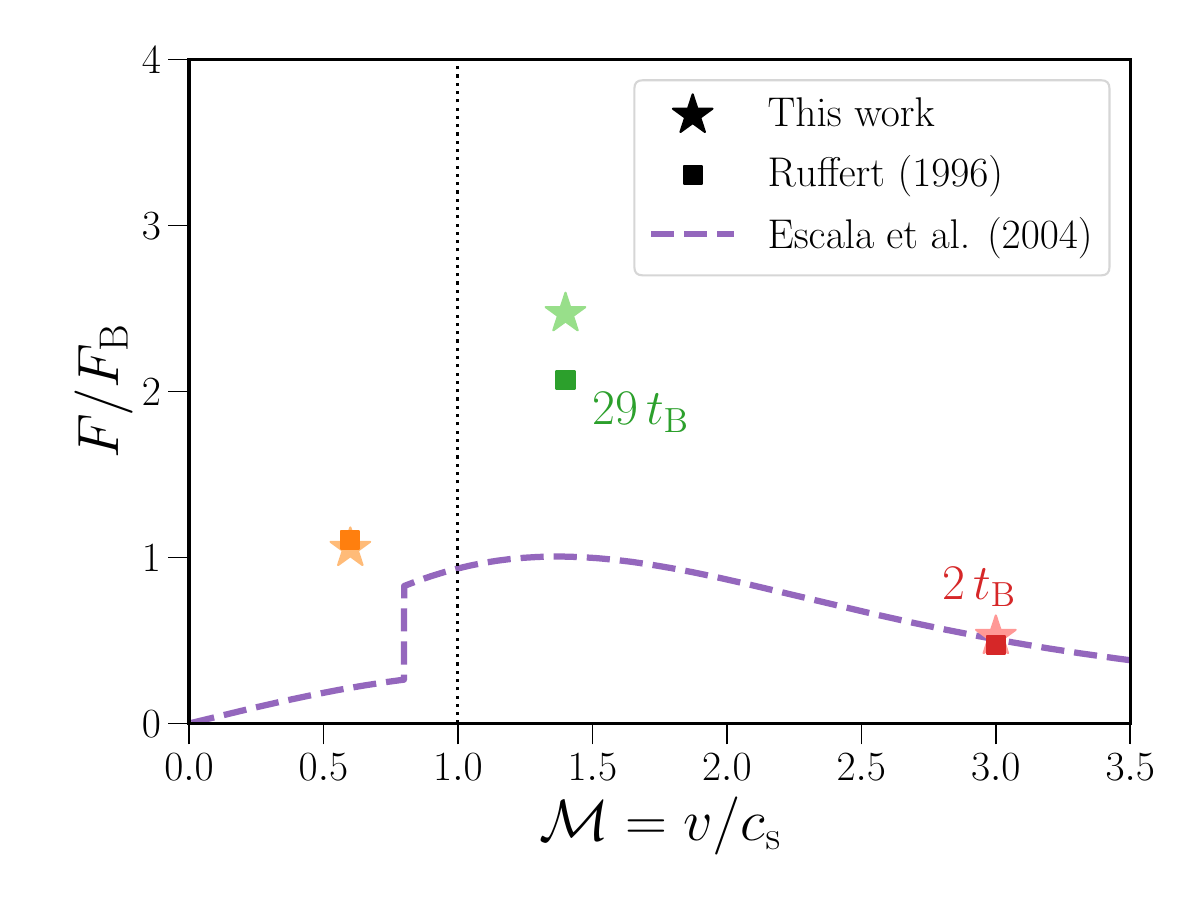}
    \caption{Same as Figure~\ref{figure:Mach_Fri} but for comparison of our results with previous works. The square and star symbols represent the results obtained by \citet{1996A&A...311..817R} and those obtained from our formula (Equation~\ref{eq:fit_form}), respectively. These symbols correspond to the results for the cases of $\mathcal{M}=0.6$ at $54\,t_{\mathrm{B}}\,(4.3\times10^5\,\mathrm{yr};\,\mathrm{orange})$, 
    $\mathcal{M}=1.4$ at $29\,t_{\mathrm{B}}\,(2.3\times10^5\,\mathrm{yr};\,\mathrm{green})$,  and $\mathcal{M}=3.0$ at $2\,t_{\mathrm{B}}\,(1.6\times10^4\,\mathrm{yr};\,\mathrm{red})$. The purple dashed line exhibits the formula provided in \citet{2004ApJ...607..765E}, which does not have explicit time dependence. }
    \label{figure:comparison}
\end{figure}

\citet{1985MNRAS.217..367S} conducted a seminal numerical study on the friction experienced by an accreting body in linear motion at a constant velocity. Building upon this, \citet{1996A&A...311..817R} performed a more comprehensive analysis using 3D hydrodynamic simulations with a nested grid technique. The latter work considered various cases where a body moving at different velocities ($\mathcal{M}=0.6,1.4,3.0,$ and $10$) accretes a nearly isothermal gas (with an adiabatic index of $\gamma=1.01$). However, the main focus of this study was to determine the accretion rates onto the body rather than to evaluate the frictonal force. 
We here compare \citet{1996A&A...311..817R} with our findings. We have carefully examined the time dependence and numerical convergence with improving the spatial resolution of the net frictional force.  

Figure~\ref{figure:comparison} illustrates the Mach number dependence of dynamical friction, similar to Figure~\ref{figure:Mach_Fri}. However, in this case, we compare our formula (Equation~\ref{eq:fit_form}; represented by stars) with the results obtained by \citet{1996A&A...311..817R} (squares). Note that \citet{1996A&A...311..817R} followed the evolution for different periods depending on the Mach numbers:  $54\,t_{\mathrm{B}}\,(4.3\times10^5\,\mathrm{yr})$ for $\mathcal{M}=0.6$, $29\,t_{\mathrm{B}}\,(2.3\times10^5\,\mathrm{yr})$ for $\mathcal{M}=1.4$, and $2\,t_{\mathrm{B}}\,(1.6\times10^4\,\mathrm{yr})$ for $\mathcal{M}=3.0$. 
To obtain our corresponding results, we substitute these times into our formula for the supersonic cases with $\mathcal{M}=1.4$ and $3.0$. There is no time dependence for the subsonic case with $\mathcal{M}=0.6$. Our results are consistent with the overall trend observed in \citet{1996A&A...311..817R}. However, there is a discrepancy of approximately 20\% between the data points at $\mathcal{M}=1.4$. This discrepancy can likely be attributed to the limited size of the computational domain used in \citet{1996A&A...311..817R}, which does not fully encompass the entire region affected by density perturbations. As a result, the friction, which should have increased steadily over time, remains constant in their study, leading to an underestimation of dynamical friction.

\subsubsection{Non-accreting body in a circular orbit}

We also compare our results with previous simulations on the dynamical
friction of a body moving in a circular orbit without accretion.
Specifically, we compare our results with \citet{2004ApJ...607..765E},
who performed SPH simulations of circularly orbiting binary BHs within a
non-singular isothermal sphere. They derived a formula to describe the velocity dependence of dynamical friction based on their simulation results.

In Figure~\ref{figure:comparison}, we show the formula provided by \citet{2004ApJ...607..765E} with the purple dashed line.  In the subsonic regime, the friction calculated with the formula of \citet{2004ApJ...607..765E} appears to be underestimated compared with our results, which might be attributed to their exclusion of gas accretion in the simulations.  In the supersonic case, their formula lacks explicit time dependence, unlike ours, possibly because the region contributing to friction is determined by the size of the core of the non-singular isothermal sphere in their simulation, as discussed in Section~\ref{sec:application}.  
However, a direct comparison is not straightforward due to other numerous differences between their and our simulations, including the treatment of accretion (included or ignored) and the assumed trajectories (linear or circular). We point out that their formula may be specialized to specific situations, which is complementary to our formula designed for general use.

For non-accreting bodies, \citet{2001MNRAS.322...67S} and \citet{2004ApJ...607..765E} argue that, around $\mathcal{M} \simeq 1$, the dynamical friction for circular orbits is suppressed compared to that for linear motion given by \citet{1999ApJ...522L..35S}, who reproduce the results of \citetalias{1999ApJ...513..252O}.  
In future studies, we intend to investigate the dynamical friction of accreting bodies that are in circular orbits.

\section{Conclusions}
\label{sec:conclusion}

We have investigated the dynamical friction acting on accreting bodies moving at a constant velocity through 3D nested-grid simulations, incorporating gas accretion with sink particles. We have systematically explored the velocity dependence of dynamical friction over a broad range of Mach numbers, $0 < \mathcal{M} < 3.0$. Furthermore, we have compared our results with the linear theory \citep{1999ApJ...513..252O}, to examine the impact of gas accretion on the dynamical friction. Our results are summarized as follows: 

\begin{itemize}
    \item 
    The net force acting on the body is the sum of gravitational force exerted by the surrounding gas and momentum flux transferred by the accreting gas (we refer to this net force as dynamical friction). 
    Through our simulations, we find that the dynamical friction is independent of the sink radius, while the gravitational force of the surrounding gas increases as the sink radius decreases. The gas between the sink and BHL radii does not contribute to the dynamical friction, because the gas within the BHL radius accelerated by the body's gravity subsequently falls onto the body, returning the momentum gained before falling. Due to the cancellation of the gravitational force within the BHL radius and the momentum flux, the dynamical friction is effectively determined by gravitational force of the gas outside the BHL radius. 

    \item 
    In the subsonic case, the dynamical friction in our simulation is larger than that of linear theory.  According to the linear theory, the region just outside the BHL radius has front-back symmetry and does not contribute to the dynamical friction. However, in our simulations, the effect of accretion breaks the front-back symmetry and introduces a new contribution from the gas in this region, leading to larger dynamical friction.

    \item 
    In the supersonic case, the linear theory \citetalias{1999ApJ...513..252O} encompasses a parameter, $r_{\mathrm{min}}$, representing the minimum radius of the gas contributing to the dynamical friction. Although $r_{\mathrm{min}}$ cannot be determined within the framework of linear theory, our simulations of accreting bodies reveal that it is physically determined. Assuming the same functional form as the linear formula \citepalias{1999ApJ...513..252O}, we find that the dynamical friction in our simulations can be reproduced with $r_{\mathrm{min}} \sim R_{\mathrm{BHL}}$ due to the cancellation of the gravitational force and the momentum flux within the BHL radius, as described above.  

    \item 
    We also find that $r_{\mathrm{min}}/\RBHL$ is a decreasing function for $1<\mathcal{M}<3.0$, which converges to become constant at $r_{\mathrm{min}}/R_{\mathrm{BHL}}=1/2$ for $\mathcal{M} \geq 3$ (Equation~\ref{eq:zeta}).  We attribute higher $r_{\mathrm{min}}/\RBHL$ for $1\leq\mathcal{M}\leq1.3$, i.e., smaller dynamical friction compared with the linear formula with $r_{\mathrm{min}}=R_{\mathrm{BHL}}/2$, to the nonlinear effect that reduces the asymmetry in the density distribution (see Figure~\ref{figure:den_sup}).  In the slightly supersonic case with $1\lesssim\mathcal{M}\lesssim1.3$, considered as transitional regime from the subsonic to supersonic regimes, the gas distribution has a similarity to the subsonic case, resulting in the largest deviation from the linear theory that predicts a distinct transition between the two regimes.

    \item 
    We provide the fitting formula of dynamical friction for accreting bodies (Equation~\ref{eq:fit_form}-\ref{eq:coefficient}), based on our simulation results. Our formula serves as an updated and improved version of the original linear formula proposed by \citetalias{1999ApJ...513..252O}. Similar to the linear formula, our revised formula can be used in a wide range of applications, such as galaxy formation simulations, where it can be employed as a subgrid model for a moving and accreting BH.       
\end{itemize}

In conclusion, we determine the velocity dependence of dynamical friction experienced by 
astronomical objects moving in a uniform medium, taking gas accretion into account. 
Note that our simulations are based on idealized setup with uniform gas density 
and that physical effects such as magnetic fields and radiation feedback are not considered in 
this study. 
Nonetheless, our results serve as a crucial initial step towards understanding the motion of astronomical objects, such as BHs, as they traverse a medium while accreting gas.

\begin{acknowledgments}
The authors thank Andr\'{e}s Escala, Ken Ohsuga, Takahiro Tanaka, Koutarou Kyutoku, Kunihito Ioka and Ryoki Matsukoba for fruitful discussions and comments. The authors also thank Kazutaka Kimura for providing useful advice on how to use SFUMATO-RT. This work was supported by JST SPRING, Grant Number JPMJSP2110. 
This research could never be accomplished without the support by Grants-in-Aid for Scientific Research (TH: 19H01934, 19KK0353, 21H00041; KS: 21K20373; TM: 18H05437, 23K03464) from the Japan Society for the Promotion of Science and Support for Pioneering Research Initiated by the Next Generation program (JPMJSP2110) from the Japan Science and Technology Agency. 
Numerical computations were carried out on Cray XC50 at Center for Computational Astrophysics, National Astronomical 
Observatory of Japan. 
\end{acknowledgments}

\bibliography{sample631}{}
\bibliographystyle{aasjournal}



\end{document}